\documentclass[12pt]{article}
\usepackage{a4wide}
\usepackage[centertags]{amsmath}
\usepackage{amssymb}
\usepackage[sort&compress,numbers]{natbib}
\usepackage{ifpdf}
\usepackage{setspace}
\usepackage{amsfonts}
\usepackage{subfigure}
\usepackage{threeparttable}
\ifpdf
\usepackage[bookmarks]{hyperref}
\else
\usepackage[hypertex]{hyperref}
\usepackage{epsfig}
\usepackage{latexsym,amsmath,amssymb,amstext}
\fi
\usepackage{enumerate}
\newcommand{\half}{\frac{1}{2}}

 \newcommand{\m}{{\mu}}

 \newcommand{\la}{{\lambda}}

 \newcommand{\g}{{\gamma}}

 \newcommand{\p}{\partial}

\newcommand{\f}{\frac}

\usepackage[dvips]{color}
\usepackage{feynmp}
\def\tr{\mathrm{tr}}
\def\Tr{\mathrm{Tr}}
\def\STr{\mathrm{STr}}

\def\half{{1\over2}}
\def\nn{\nonumber\\}

\def\[{\left[}
\def\]{\right]}
\def\({\left(}
\def\){\right)}

\newcommand{\wt}{\widetilde}

\def\={\stackrel{\bullet}{=}}
\def\nn{\notag\\}

\def\Tr{\mathrm{Tr}}

\def\half{{1\over2}}

\def\[{\left[}
\def\]{\right]}
\def\({\left(}
\def\){\right)}

\def\cD{{\cal D}}

\def\cL{{\cal L}}
\def\cN{{\cal N}}
\def\cO{{\cal O}}

\def\bS{{\mathbf S}}

\def\bR{{\mathbf R}}
\def\bZ{{\mathbf Z}}

\def \be {\begin{equation}}
\def \ee {\end{equation}}
\def \bea {\begin{eqnarray}}
\def \eea {\end{eqnarray}}
\def \beal#1 {\begin{align}#1\end{align}}
\def \nn {\notag\\}
\def\sla#1{\not\!\!#1}

\def\bra#1{\langle #1 |}
\def\ket#1{|#1 \rangle}
\def\aver#1{\left\langle #1 \right\rangle}

\def\beal#1{\begin{align}#1\end{align}}

\begin{document}
\setlength{\topmargin}{-.1in} 
\setlength{\textheight}{9.0in}
\makeatletter
\renewcommand{\theequation}{%
\thesection.\arabic{equation}}
\@addtoreset{equation}{section}
\makeatother

\begin{titlepage}
\vspace{-3cm}
\title{
\begin{flushright}
\normalsize{ TIFR/TH/12-30\\
July 2012}
\end{flushright}
       \vspace{1.5cm}
Supersymmetric  Chern-Simons Theories \\
with Vector Matter
       \vspace{1.5cm}
}
\author{
\!\!\!\!\!\!\!\! 
Sachin Jain$^{a),*}$, 
Sandip P. Trivedi$^{a),\dagger}$,
Spenta R. Wadia$^{a),b),\ddagger}$, 
Shuichi Yokoyama$^{a),\star}$
\\[25pt] 
\!\!\!\!\!\!\!\!{\it \normalsize $^{a)}$Department of Theoretical Physics, Tata Institute of Fundamental Research,}\\
\!\!\!\!\!\!\!\!{\it \normalsize Homi Bhabha Road, Mumbai 400005, India}\\
\!\!\!\!\!\!\!\!{\it \normalsize $^{b)}$International Centre for Theoretical Sciences, Tata Institute of Fundamental Research,} \\
\!\!\!\!\!\!\!\!{\it \normalsize TIFR Centre Building, Indian Institute of Science, Bangalore 560004, India}
\\[20pt]
\!\!\!\!\!\!\!\!{\small \tt E-mail: $^*$sachin, ${}^\dagger$sandip, ${}^\ddagger$wadia, ${}^\star$yokoyama(at)theory.tifr.res.in}
}

\date{}

\maketitle

\thispagestyle{empty}

\vspace{.5cm}

\begin{abstract}
\vspace{0.3cm}
\normalsize
In this paper we discuss $SU(N)$ Chern-Simons theories at level $k$ with both fermionic and bosonic vector matter. In particular we present an exact calculation of the free energy of the $\cN=2$ supersymmetric model (with one chiral field) for all values of the 't Hooft coupling in the large $N$ limit. This is done by using a generalization of the standard Hubbard-Stratanovich method because the SUSY model contains higher order polynomial interactions. 
\end{abstract}

\end{titlepage}

\tableofcontents

\section{Introduction}
\label{intro}

Three dimensional conformal field theories are of great interest in the context of AdS/CFT and
dS/CFT correspondence \cite{Maldacena:1997re}, \cite{Strominger:2001pn}. 
An important feature is the duality \cite{Klebanov:2002ja} between  
the singlet sector of a vector model and a higher spin gravity theory \cite{Vasiliev:1990en,Vasiliev:2003ev,Bekaert:2005vh}. 
A typical example is the correspondence between the singlet sector of 
$O(N)\, (U(N))$ free/critical vector model and Vasiliev theory defined on AdS$_4$ with suitable  boundary conditions \cite{Klebanov:2002ja,Sezgin:2002rt,Sezgin:2003pt}.
Evidence for this conjecture was provided by computing the correlation functions of the higher spin gauge theory \cite{Giombi:2010vg,Giombi:2011rz,Giombi:2011ya}.

Subsequently vector matter was gauged along with the addition of a conformally invariant Chern-Simons term \cite{Giombi:2011kc}, 
\cite{Aharony:2011jz}. 
This ensures that all operators and states of the theory are gauge singlets. The corresponding dual higher spin theory also
violates parity. Supersymmetric extensions of parity violating Vasiliev type higher spin theories have been recently constructed in
 \cite{Chang:2012kt}.
Related works are \cite{Shenker:2011zf,Hellerman:2012}.


There are three main results in \cite{Giombi:2011kc}, where $SU(N)$ level $k$ Chern-Simons theory with a single fermion was discussed in the large $N$ limit: 
i) calculation of the thermal 2-point functions and consequently a formula for the free energy as a function of the 't Hooft coupling and the temperature,
ii) identification of the operator spectrum at large $N$ with an infinite set of single trace gauge invariant currents of integer spin $s$ and dimension $ s+1 + \cO(1/N)$, 
iii) determination of the equation of the divergence of these currents,  
which is schematically given by
\be
\partial \cdot j = {1\over N} jj + {1\over N^2} jjj.
\label{conse}
\ee
Conservation and partial conservation of higher spin currents have been used to derive very general results about current correlation functions in \cite{Maldacena:2011jn,Maldacena:2012sf}.

The derivation of the thermodynamic results rested on the fact that i) the field theory was defined on $\bR^2\times \bS^1$ where one could choose the `light cone' gauge%
\footnote{This gauge is explained in the main text, \eqref{gauge}.} 
in the non-thermal directions and ii) on dimensional regularization working in $2+1-\epsilon$ dimensions. The choice of the gauge enables a representation of the path-integral in terms of bilocal fields which at the large $N$ saddle point is the 2-point correlation function that solves the large $N$ Schwinger-Dyson equations.

This paper is a natural continuation of \cite{Giombi:2011kc} by including not only a fundamental fermion but also a fundamental scalar (Chern-Simons-scalar theories were considered in \cite{Aharony:2011jz}). One of our main motivations is to study  the ${\cal N}=2$ supersymmetric (SUSY) theory with one chiral field in the fundamental representation, which arises for special values of the parameters in this class of theories. 
In this case, the conformal symmetry is exactly preserved \cite{Gaiotto:2007qi}.

The new features in the present study are the double trace Yukawa coupling and the triple trace (sextic) coupling that appear in the Lagrangian. 
To deal with them we develop a general method  for carrying out path integrals in the large $N$ limit that goes beyond the standard Hubbard-Stratanovich method \cite{Hubbard:1959ub}, \cite{Stratonovich:1959} that applies to quartic interactions. This 
general method  is in fact valid for any polynomial action involving local as well as bilocal fields and uses two auxiliary conjugate fields of a functional Fourier transform. At the large $N$ saddle point these conjugate fields have the interpretation of the 2-point function (at large $N$) and the corresponding self-energy. The saddle point equations can be shown to be the Schwinger-Dyson equations at large $N$.

%

Our exact result for $\cN=2$ SU($N$) level $k$ Chern-Simons theory with a fundamental chiral multiplet, 
on $\bR^2$  at temperature $T$,  in the 't Hooft limit is the following. 
The self-energies of the scalar and fermion are
\beal{
\Sigma_B(p) &= c T^2, \quad \Sigma_F(p) = f p_s +i \left({T^2 c \over p_s^2} -f^2 \right) \gamma^+ p_+,
}
where $\gamma^+={1\over \sqrt2}(\gamma_1+i\gamma_2),\, p_+={1\over \sqrt2}(p_1-ip_2),\, p_s = \sqrt{p_1^2+p_2^2}$ ($p_1, p_2$ are the components of momentum in $\bR^2$) and 
\beal{
f &= \frac{2\lambda}{\beta p_s}
\biggl(\log(2\cosh(\frac{\sqrt{(\beta p_s)^2 + c}}{2})) - \log({2\sinh(\frac{\sqrt c}{2})})\biggr), \quad 
\sqrt c = 2 |\lambda|  \ln( \coth\frac{\sqrt{c}}{2}). 
}
The free energy density is 
\beal{
F&= -{NT^3 \over 6\pi}  \biggl({ \sqrt c^3 \over |\lambda| } + 
6 \int _{\sqrt c}^\infty dy y \log ( \coth {y \over 2} ) \biggr).
}
Note that the thermal mass $c$ remains finite for all finite values of $\lambda$ and only 
diverges as $c \sim \ln|\lambda|$ as $\lambda
\rightarrow \infty$. Similarly, the free energy is also finite and non-vanishing for all finite values of $\lambda$. Its magnitude 
 decreases monotonically from the free field value  as $|\lambda|$ is increased and vanishes when $|\lambda|\rightarrow \infty$.
These results are consistent with the conformal theory existing for all values of $\lambda$.  
This behavior is also different from that in \cite{Giombi:2011kc}, where it was shown that  for a fundamental fermion coupled to the CS gauge field
 the  free energy vanishes at $|\lambda|=1$.

The rest of this paper is organized as follows. 
In Section \ref{CSM}, we present the exact solution of SU($N$) level $k$ Chern-Simon theory coupling to  fundamental scalar and fermion
 matter   in the 't Hooft limit on $\bR^3$. 
 In  particular, in Section \ref{rewritingbysinglet} we present the
 general method for doing the large N path integral in the presence of higher order  interactions
using a functional Fourier transform.    
 In Section \ref{fta}, we solve the same system at  finite temperature by analyzing it on $\bR^2 \times \bS^1$. 
Most of the analysis  in these sections is  carried out for a general theory consisting of one fermion and one scalar in the fundamental representation of the gauge group with  varying values of the 't Hooft coupling, the Yukawa and sextic couplings.
In Section \ref{susyapply} we specialise to the ${\cal N}=2$ case and present the results explicitly for that theory. 
 Section \ref{discussion}  is devoted to  discussion and future directions. 
Appendices contain the conventions, technical details and consistency checks of the results in the main text.

\bigskip

\paragraph{Note added:} 
While  preparing this manuscript we became aware of the work of O.Aharony, G.Gur-Ari and R.Yacoby \cite{Aharony:2012nh}. We thank the authors for informing us in advance. 
We also mention that the thermal partition function involving only the CS plus scalar fields has also been calculated in 
\cite{Oferb:2012}.

\section{Chern-Simons theory with fundamental matter in the 't Hooft limit}
\label{CSM}
\subsection{Action}

In this section we present the exact solution of Chern-Simons theory coupling to fundamental bosons and fermions with conformal symmetry 
in the 't Hooft limit. For this purpose, we specifically study $SU(N)$ Chern-Simons gauge theory with level $k$ coupled to a single 
fundamental boson $\phi$ and fermion $\psi$ on $\bR^3$. The action is given by
\beal{
S  &= \int d^3 x  \biggl[i \varepsilon^{\mu\nu\rho} {k \over 4 \pi}
\Tr( A_\mu\partial_\nu A_\rho -{2 i\over3}  A_\mu A_\nu A_\rho)
+ D_\mu \bar \phi D^\mu\phi + \bar\psi \gamma^\mu D_\mu \psi \nn
&
+ \lambda_4 (\bar\psi \psi) (\bar\phi\phi)
+ \lambda_4' (\bar\psi\phi)( \bar\phi \psi)
+ \lambda_4'' \left((\bar\psi \phi)( \bar \psi \phi ) +(\bar \phi \psi)( \bar \phi \psi )
\right)+ \lambda_6 (\bar\phi\phi)^3\biggl].
\label{generalaction}
}
Here Levi-Civita symbol $\varepsilon^{\mu\nu\rho}$ is normalized as 
$\varepsilon_{123}=\varepsilon^{123}=1$. 
The covariant derivative acts on fields as follows (See Appendix \ref{convention} for  relevant conventions).
\beal{
D_{\m} \phi &= \p_{\m} \phi-i A^a_{\m} T^a \phi,\quad D_{\m} \bar{\phi} = \p_{\m} \bar{\phi}+i A^a_{\m} \bar{\phi} T^a,  \\
D_{\m} \psi &= \p_{\m} \psi-i A^a_{\m} T^a \psi,\quad
D_{\m}  \bar{\psi} = \p_{\m}  \bar{\psi}+i A^a_{\m} \bar{\psi}T^a.
}

The 't Hooft limit is defined by 
\be
N, k \to \infty \quad \mbox{with $\lambda:= {N\over k}$ fixed.} 
\label{thooftlimit}
\ee
Accordingly we define the coupling constants by
\be
\lambda_4={x_4 \over \kappa}, \quad 
\lambda'_4={x'_4 \over 2\kappa}, \quad 
\lambda''_4={x''_4 \over 4\kappa}, \quad 
\lambda_6={x_6 \over (2 \kappa)^2},
\label{coefficients}
\ee
where we have set  $\kappa = { k \over 4 \pi}$. 
We keep the coefficients $x_4, x_4', x_4'', x_6$ of order one in the 't Hooft limit.
Note in advance that the relevant coefficients to the self energy and the free energy are $x_4, x_6$ and the others do not matter in the 't Hooft limit.

\subsection{Exact effective action}

\subsubsection{Gauge fixing}

The first important step to solve the gauge theory is gauge fixing. 
Since the gauge theory is being defined in $\bR^2\times \bR^1$ or $\bR^2\times \bS^1$ we fix the gauge degrees of freedom by using the following method. 
We first work with $\bR^{1,1}\times \bR^1$ or $\bR^{1,1} \times \bS^1$.
$\bR^{1,1}$ is a Lorentzian space-time. 
Here we fix the light-cone gauge
\be
\label{gauge}
 A_{-} :={1\over \sqrt2}(A_1+A_2)=0.
\ee
Note that this gauge preserves Poincare symmetry in $\bR^{1,1}$. 
In this gauge the Chern-Simons gauge interaction vanishes and 
the gauge field component $A_{+}={1\over \sqrt2}(A_1-A_2) $ is linear in the action.
The action reduces to 
\beal{
~\label{LCaction1}
S&=\int d^3x \biggl[ \f{k}{2 \pi} A_{+}^a \p_{-} A_{3}^a  
- \bar{\phi}(\partial_3^2 + 2 \p_{+} \p_{-} ) \phi + \bar{\psi} \gamma^\mu \p_{\mu} \psi +\nonumber \\
&+ A_{ +}^a J_{-}^a + A_{3}^a J_{3}^a
+A_{3}^a A_{3}^b \bar\phi T^a T^b \phi \nn
&+ \lambda_6 (\bar\phi\phi)^3 + \lambda_4 (\bar\psi \psi) (\bar\phi\phi)
+ \lambda_4' (\bar\psi\phi)( \bar\phi \psi)
+ \lambda_4'' \left((\bar\psi \phi)( \bar \psi \phi ) +(\bar \phi \psi)( \bar \phi \psi )\right)\biggl]
}
where $J_{\mu}^a=i (\bar{\phi} T^a \p_{\m} \phi-(\p_{\m} \bar{\phi}) T^a \phi)-i \bar{\psi} \g_{\m} T^a \psi$. 
Here and hereafter we omit the total derivative terms by appropriate boundary conditions. 
Since $A_+$ is linear, we can integrate it out leading to the following constraint. 
\be
\f{k}{2 \pi} \partial_- A_{3}^a + J_{-}^a =0.
\ee
To solve this we Fourier transform to momentum space
\be
A_{3}^a(p) =  \f{2 \pi i}{kp_- } J_{-}^a(p), 
\ee
where $p_-$ is understood using the principle value prescription $p_- \rightarrow p_-+i\epsilon$ and
\beal{
J_{-}^a(p) &=- \half \int {d^3q\over (2\pi)^3} ( (2q+ p)_- \bar\phi(-q)T^a \phi (q+p)
+ i \bar\psi(-q)\gamma_-T^a \psi (q+p)).
}
Plugging this back we obtain an action with the gauge field integrated out, 
which is denoted by $S_1$. 
At this point we analytically continue back from $\bR^{1,1}$ to $\bR^2$. 
In terms of matter content, the action $S_1$ can be divided into three parts. 
\be
S_1= S_B + S_F +S_{BF}. 
\label{integratedaction}
\ee
The first part $S_B$ consists only of bosonic fields and is given by
\beal{
S_B&= \int \f{d^3q}{(2 \pi)^3}    (q_s^2 +q_3^2) \bar{\phi}(-q)\phi(q) +
N \int  \f{d^3P}{(2 \pi)^3} \f{d^3q_1}{(2 \pi)^3} \f{d^3q_2}{(2 \pi)^3} 
 C_1(P,q_1,q_2)\chi(P,q_1) \chi(-P,q_2) \nn
&+ N \int \f{d^3P_1}{(2 \pi)^3} \f{d^3P_2}{(2 \pi)^3}  \f{d^3q_1}{(2 \pi)^3} \f{d^3q_2}{(2 \pi)^3}\f{d^3q_3}{(2 \pi)^3} C_2(P_1,P_2,q_1,q_2,q_3)
\chi(P_1,q_1)\chi(P_2,q_2)\chi(-P_1-P_2,q_3).
\label{sb}
}
Here $\chi(P,q)$ is a bilinear of the scalar fields with the gauge index  contracted
\beal{
\chi(P,q) &= {1 \over N} \bar\phi({P \over 2}-q) \phi({P \over 2}+q),
\label{chi}
}
where $P$ represents the total momentum of the bilinear and $q$ relative momentum. The coefficients $C_1(P,q_1,q_2), C_2(P_1,P_2,q_1,q_2,q_3)$ are 
\beal{
C_1(P,q_1,q_2)&={2 \pi i N \over k} \f{(-P+q_1+ q_2)_{3} (P+q_1+q_2)_{-}}{(q_1-q_2)_{-}}, \\
C_2(P_1,P_2,q_1,q_2,q_3)&=
{4 \pi^2 N^2 \over k^2} \f{( P_1 - P_2+2q_1+2q_2)_{-} (P_1+2P_2+ 2q_2 + 2q_3)_{-}}{(P_1+P_2+ 2q_1-2q_2)_{-} (P_1-2q_2+2q_3)_{-}} + N^2 \lambda_6.
}
The second part $S_F$ is constituted only from the fermionic fields. The action in the leading order of large $N$ limit is given by
\beal{
S_F=& i \int  \f{d^3 q}{(2\pi)^3} \bar{\psi}(-q) \gamma^\mu q_\mu \psi(q)  \nn
&+ \int \f{d^3 p}{(2 \pi)^3}\f{d^3 q}{(2 \pi)^3} \f{d^3 r}{(2 \pi)^3} {-2\pi i \over k p_-} \bar\psi(-q)\gamma_-T^a \psi(q+p) 
\bar\psi(-r)\gamma_3 T^a \psi(r+p) \nn
=& i \int  \f{d^3 q}{(2\pi)^3} \bar{\psi}(-q) \gamma^\mu q_\mu \psi(q)  
+N \int \f{d^3P}{(2 \pi)^3} \f{d^3q_1}{(2 \pi)^3} \f{d^3q_2}{(2 \pi)^3}~
\f{8 \pi iN}{ k (q_1-q_2)_{-}} \xi_-(P,q_1) \xi_I(-P,q_2),
\label{sf}
}
where we used \eqref{tata} and $\xi_I(P,q), \xi_-(P,q)$ are bilinear of the fermionic field with gauge and spinor indices contracted.
\beal{
\xi_I(P,q) = \frac{1}{2N}\bar{\psi}(\f{P}{2}-q) \psi(\f{P}{2}+q), \quad
\xi_-(P,q) = \frac{1}{2N}\bar{\psi}(\f{P}{2}-q) \gamma_-\psi(\f{P}{2}+q).
\label{xi}
}
The last term $S_{BF}$ involves both the bosonic and fermionic fields.
\beal{
S_{BF}&=N  \int \f{d^3P}{(2 \pi)^3} \f{d^3q_1}{(2 \pi)^3} \f{d^3q_2}{(2 \pi)^3}
2 \lambda_4 N \chi(P,q_1) \xi_I(-P,q_2) + \Delta S_{BF}(\eta,\bar{\eta}),
\label{sbf}
}
where $\Delta S_{BF}(\eta,\bar{\eta})$ includes fermionic bilinear operators $\eta(P,q), \bar{\eta}(P,q)$ defined by
\beal{
\eta (P,q) = \frac{1}{N}\bar{\phi}(\f{P}{2}-q)\psi(\f{P}{2}+q), \quad
\bar{\eta} (P,q) = \frac{1}{N}\bar{\psi}(\f{P}{2}-q) \phi(\f{P}{2}+q).
}
The precise formulas are available in Appendix \ref{sbf}.
For the large $N$ saddle points considered in this paper, 
the classical values of these fields $\eta(P,q), \bar{\eta}(P,q)$ vanish.

\subsubsection{Method of functional Fourier transform for the large $N$ limit}
\label{rewritingbysinglet}

In the previous subsection we have integrated out the gauge fields from the action \eqref{generalaction} and obtained the action \eqref{integratedaction} containing bosonic and fermionic bilocal fields. 
This leads to the complete elimination of the gauge field and 
the action is written entirely in terms of bilocal fields:
\be
\sum_{m=1}^N \bar\psi_m^\alpha(x_1) \psi^m{}^\beta(x_2), \quad
\sum_{m=1}^N \bar\phi_m(x_1) \phi^m{}(x_2), \quad
\sum_{m=1}^N \bar\psi_m^\alpha(x_1) \phi^m{}(x_2), \quad
\sum_{m=1}^N \bar\phi_m(x_1) \psi^m{}^\alpha(x_2), \quad
\ee
which are single trace operators, and kinetic parts like $\bar\psi \gamma^\mu\partial_\mu \psi $, $\partial^\mu \bar\phi \partial_\mu \phi$, and some local terms. 
The appearance of bilocal fermionic and bosonic operators is a reflection of the fact that this theory contains higher spin fields.

\bigskip

We now develop a general method for carrying out path integrals in the large $N$ limit for arbitrary polynomial interaction.%
\footnote{
Our method is an adaptation of a method developed for multi-trace unitary matrix model, see Section 5 of \cite{AlvarezGaume:2006jg}. 
}
It is best to illustrate the main idea of the method in a notationally simple example of bosonic fields. 
\beal{
S_{ex} &= \int d^3 x \partial_\mu \bar\phi \partial^\mu \phi 
+ \int d^3 x_1d^3 x_2d^3 x_3d^3 x_4 K_4(x_1, x_2, x_3, x_4) (\bar\phi(x_1) \phi(x_2)) (\bar\phi(x_3) \phi(x_4)) \nn
&\quad + \int d^3 x_1d^3 x_2 \cdots d^3 x_6 K_6(x_1, x_2, \cdots , x_6)(\bar\phi(x_1) \phi(x_2)) (\bar\phi(x_3) \phi(x_4)) (\bar\phi(x_5) \phi(x_6)) + \cdots \nn
&=: \int d^3 x \partial_\mu \bar\phi \partial^\mu \phi + V(\bar\phi \phi).
}
Introduce two conjugate fields $\alpha, \mu$ such that
\beal{
1 &=\int \cD\alpha \delta(\alpha(x_1, x_2) - {1\over N} \bar\phi(x_1)\phi(x_2)) \\
&=\int \cD\alpha \cD\mu \exp\biggl[{i \int d^3x_1 d^3 x_2 \mu(x_1, x_2) (\alpha(x_1, x_2) - {1\over N} \bar\phi(x_1)\phi(x_2))}\biggl].
}
We insert this identity into the partition function and integrate out the scalar fields $\phi$ and $\bar\phi$
\beal{
Z_{ex} &= \int \cD\bar\phi \cD\phi \, e^{-S_{ex}} \\
&=\int \cD\alpha \cD\mu \exp\biggl[{i \int \mu \cdot \alpha -V(N \alpha)}\biggl]\nn
&\qquad \times \int \cD\bar\phi \cD\phi \,  \exp\biggl[{i \int d^3x_1 d^3 x_2 \left(\partial_\mu\bar\phi \partial^\mu\phi - \mu(x_1, x_2) {1 \over N} \bar\phi(x_1)\phi(x_2)\right)}\biggl] \\
&=\int \cD\alpha \cD\mu \exp\biggl[{N \left(i \int \wt \mu \cdot \alpha - \wt V( \alpha) -N \tr \log(-\partial^2 + i \wt \mu) \right)}\biggl],
}
where we set%
\footnote{
If $V(\alpha)$ is quadratic in $\alpha$, the path integral with respect to $\alpha$ in $Z$ reduces to Gaussian integral. This is the 
Hubbard-Stratanovich method and one
 is left with a Gaussian integral over $\mu$.
If $V(\alpha)$ is not quadratic, we have to work with both $\alpha$ and $\wt \mu$.
}
\be
V(N \alpha) = N \wt V(\alpha), \quad \mu(x_1, x_2) = N \wt \mu(x_1, x_2).
\ee
After appropriate scaling by $N$ the action is
\be
S_{ex} = N \biggl(  \tr \log (-\partial^2 + i\wt \mu)  + \wt V(\alpha) - i  \int \wt \mu \cdot \alpha \biggl)
\ee
and the large $N$ saddle point equations are
\beal{
&{\delta \wt V \over \delta \alpha(x,y)}  + i\wt \mu(y,x)=0,
\label{a}\\
&i\alpha(x, y) - {\delta \over \delta \wt\mu(x,y)} \tr\log(-\partial^2+i\wt\mu)=0.
\label{b}
}
From \eqref{a}, \eqref{b} we  obtain 
\begin{eqnarray}
\alpha(x, y) &=& \bra{x}{1 \over -\partial^2 + i \wt\mu} \ket{y}.\label{alphapropa}
\end{eqnarray}
We note that at the saddle point
$\alpha$ is the propagator and $i\wt\mu\equiv \Sigma$ is the self-energy to leading order in large $N$. This equation can be written symbolically as
\be
( -\partial^2 + i \wt\mu ) \alpha = {\bf 1},
\ee
where the rhs is the identity operator with matrix elements $ \bra{x}{\bf 1}\ket{y} = \delta(x-y)$.

By using \eqref{a} we find
\be
\left( -\partial^2 -{\delta \wt V \over \delta \alpha(x,y)}  \right) \alpha = {\bf 1}.
\ee
These are precisely the Schwinger-Dyson equations after using large $N$ factorization. 
\be
\int \cD\bar\phi\cD\phi {\delta \over \delta \bar\phi(x)_m}\left[\bar\phi(y)_m e^{-S_{ex}}
\right] =0.
\ee
This can be explicitly checked and it emphasizes the intimate relation between the existence of the saddle point equations and large N factorization.

The free energy $F_{ex}=- \log Z_{ex}$ evaluated at the saddle point is
\be
F_{ex} = N \biggl(\tr \log (-\partial^2 + i \aver{\wt \mu}) + \wt V(\aver{\alpha}) - i  \int \aver{\wt \mu} \cdot \aver{\alpha} \biggl),
\ee
where $\aver{\alpha}, \aver{\wt\mu}$ are solutions of the saddle point equations. The above discussion could have been done as well in momentum space.

It is important to note that the above discussion implicitly assumes the presence of an infrared cutoff and that the large N limit is taken before this cutoff is removed. In a sense $N$ is the largest parameter in the theory. This method would fail if the infrared cutoff (volume of space) itself was of $\cO(N)$.

\bigskip

Let us apply this method to our case for which the action \eqref{integratedaction} is given by
\beal{
S_1 &=
 \int \f{d^3q}{(2 \pi)^3}    (q_s^2 +q_3^2) \bar{\phi}(-q)\phi(q) +
 i \int  \f{d^3 q}{(2\pi)^3} \bar{\psi}(-q) \gamma^\mu q_\mu \psi(q)  +N\wt V(\chi, \xi_I, \xi_-, \eta, \bar\eta),
}
where
\beal{
\wt V(\chi, \xi_I, \xi_-, \eta, \bar\eta) 
&=\int  \f{d^3P}{(2 \pi)^3} \f{d^3q_1}{(2 \pi)^3} \f{d^3q_2}{(2 \pi)^3} 
 C_1(P,q_1,q_2)\chi(P,q_1) \chi(-P,q_2) \nn
&+  \int \f{d^3P_1}{(2 \pi)^3} \f{d^3P_2}{(2 \pi)^3}  \f{d^3q_1}{(2 \pi)^3} \f{d^3q_2}{(2 \pi)^3}\f{d^3q_3}{(2 \pi)^3} C_2(P_1,P_2,q_1,q_2,q_3)
\chi(P_1,q_1)\chi(P_2,q_2)\chi(-P_1-P_2,q_3) \nn
&+\int \f{d^3P}{(2 \pi)^3} \f{d^3q_1}{(2 \pi)^3} \f{d^3q_2}{(2 \pi)^3}~
\f{8 \pi i N}{ k (q_1-q_2)_{-}} \xi_{-}(P,q_1) \xi_{I}(-P,q_2) \nn
&+ \int \f{d^3P}{(2 \pi)^3} \f{d^3q_1}{(2 \pi)^3} \f{d^3q_2}{(2 \pi)^3}
2 \lambda_4 N \chi(P,q_1) \xi_I(-P,q_2) +{1\over N} \Delta S_{BF}(\eta,{\bar{\eta}}).
\label{s1'}
}
The partition function can be written as
\beal{
Z &= \int \cD\bar\psi \cD\psi\cD\bar\phi \cD\phi \, \exp\biggl[-\bigl( \int \f{d^3q}{(2 \pi)^3}    (q_s^2 +q_3^2) \bar{\phi}(-q)\phi(q) \nn
&\qquad\qquad +i \int  \f{d^3 q}{(2\pi)^3} \bar{\psi}(-q) \gamma^\mu q_\mu \psi(q) 
+N \wt V(\chi, \xi_I, \xi_-, \eta, \bar\eta)\bigl)\biggl]. \label{partitionf}
}
Now as illustrated above we introduce pairs of conjugate fields 
$
(\alpha_B, \mu_B)$,$ (\alpha_{F,I}, \mu_{F,I})$, $(\alpha_{F,-}, \mu_{F,+})$, $(\alpha_\eta, \mu_\eta)$
$ (\alpha_{\bar\eta}, \mu_{\bar\eta}) $
such that
\beal{
1 &=  \int \cD {\alpha_B} \cD {\alpha_{F,I}} \cD {\alpha_{F,-}} \cD {\alpha_\eta} \cD {\alpha_{\bar\eta}} \nn
&\qquad  \delta\left({\alpha_B} - {1 \over N} \bar\phi \phi \right)
  \delta\left( {\alpha_{F,I}} -{1 \over 2N} \bar\psi \psi \right )  
 \delta\left( {\alpha_{F,-}} -{1 \over 2N} \bar\psi \gamma_- \psi \right ) 
 \delta\({\alpha_{\eta}}-{1 \over N} \bar\phi \psi\) 
 \delta\({\alpha_{\bar\eta}}-{1 \over N} \bar\psi \phi\)  \nn
&=  \int  \cD {\alpha_B} \cD {\alpha_{F,I}} \cD {\alpha_{F,-}} \cD {\alpha_\eta} \cD {\alpha_{\bar\eta}}  \cD {\mu_B} \cD {\mu_{F,I}} \cD {\mu_{F,-}} \cD {\mu_\eta} \cD {\mu_{\bar\eta}}  \nn
&\; ~ \exp \biggl[ i\int \Bigl(\mu_B \cdot (\alpha_B- {1 \over N} \bar\phi \phi) +2 \mu_{F,+}\cdot (\alpha_{F,-}-{1 \over 2N} \bar\psi \gamma_- \psi)
+2 \mu_{F,I} \cdot (\alpha_{F,I}-{1 \over 2N} \bar\psi \psi)   \nn
& \qquad \qquad 
+ \mu_\eta \cdot (\alpha_\eta -{1 \over N} \bar\phi \psi) 
+ \mu_{\bar\eta} \cdot (\alpha_{\bar\eta}-{1 \over N} \bar\psi \phi) \Bigl) \biggr],
\label{unitinsertion}
}
where the factor $2$ is inserted in front of $\mu_F$ due to the fact that $\xi_I, \xi_-$ in \eqref{xi} have the coefficient $\half$.  
Inserting this   identity into (\ref{partitionf}) we get
\beal{
Z &=\int \cD\boldsymbol\alpha \cD\boldsymbol\mu \exp\biggl[{i \int \boldsymbol\mu \cdot \boldsymbol\alpha - N \wt V(\alpha_B, \alpha_{F,I}, \alpha_{F,-}, \alpha_\eta, \alpha_{\bar\eta})}\biggl] \times \int \cD\bar\psi \cD\psi\cD\bar\phi \cD\phi \, 
 e^{-S_{free}},
}
where
\beal{
\int \cD\boldsymbol\alpha \cD\boldsymbol\mu  \exp\biggl[i \int \boldsymbol\mu \cdot \boldsymbol\alpha \biggl] &=  \int  \cD {\alpha_B} \cD {\alpha_{F,I}} \cD {\alpha_{F,-}} \cD {\alpha_\eta} \cD {\alpha_{\bar\eta}}  \cD {\mu_B} \cD {\mu_{F,I}} \cD {\mu_{F,-}} \cD {\mu_\eta} \cD {\mu_{\bar\eta}}  \nn
&\; ~ \exp \biggl( i\int \left(\mu_B \cdot \alpha_B +2 \mu_{F,+}\cdot \alpha_{F,-}
+2 \mu_{F,I} \cdot \alpha_{F,I} + \mu_\eta \cdot \alpha_\eta  + \mu_{\bar\eta} \cdot \alpha_{\bar\eta} \right) \biggr),\label{mualpha}
}
and $S_{free}$ is given by
\beal{
S_{free} &=\int \f{d^3q}{(2 \pi)^3}  \biggl(  (q_s^2 +q_3^2) \bar{\phi}(-q)\phi(q) 
+ i \bar{\psi}(-q) \gamma^\mu q_\mu  \psi(q) +i \int  \f{d^3P}{(2 \pi)^3} \bigl(-\mu_B (P,q) \chi(-P,q) \nn
& -2 \mu_{F,+}(P,q) \xi_{-}(-P,q)-2 \mu_{F,I}(P,q) \xi_{I}(-P,q)-\mu_\eta (P,q) \eta(-P,q)-\mu_{\bar\eta} (P,q) {\bar\eta}(-P,q) \bigl)
\biggr) \nn
&=:\int \f{d^3P}{(2 \pi)^3} \f{d^3q}{(2 \pi)^3} \bigl(\bar\phi({P\over 2}-q), \bar\psi({P\over 2}-q)\bigl)
Q(\boldsymbol\mu) 
\left(\begin{array}{c} \phi({P\over2}+q) \\ \psi({P\over2}+q) \end{array}\right)\label{Sfree},
}
which is the quadratic in elementary fields. Note that in (\ref{Sfree}) we have used
\bea 
Q(\boldsymbol\mu) 
= \left(
\begin{array}{cc}
(q_s^2+q_3^2) \delta(P) +{i\wt\mu_B} & {i\wt\mu_\eta} \\
{i\wt\mu_{\bar\eta}} & i \gamma^\mu q_\mu\delta(P) +  \gamma^+ i\wt\mu_{F,+} +i\wt\mu_{F,I} 
\end{array}\right),
\eea
\begin{eqnarray}
 \mu_B &=& N \wt \mu_B,~~ \mu_{F_{I}}= N \wt \mu_{F_{I}},~~\mu_{F_{+}}= N \wt \mu_{F_{+}},~~\mu_{\eta}= N \wt \mu_{\eta}, ~~\mu_{\bar {\eta}}= N \wt \mu_{\bar {\eta}}.
 \end{eqnarray}
By performing path integral for the elementary fields $\phi, \bar\phi, \psi, \bar\psi$ we obtain\footnote{ Since (\ref{Sfree})
 is quadratic in fields $\phi, \bar\phi, \psi, \bar\psi$, one can easily do the path integral by using Gaussian integrals.} 
\beal{
Z &=\int \cD\boldsymbol\alpha \cD\boldsymbol\mu \exp\biggl[{N\left(i \int \wt{\boldsymbol\mu} \cdot \boldsymbol\alpha - \wt V(\alpha_B, \alpha_{F,I}, \alpha_{F,-}, \alpha_\eta, \alpha_{\bar\eta} ) - \STr \log Q(\boldsymbol\mu)  \right)}\biggl] .  
}
Therefore we obtain the effective action as
\be
S_{eff} =N\left(\STr \log Q(\boldsymbol\mu)   + \wt V(\alpha_B, \alpha_{F,I}, \alpha_{F,-}, \alpha_\eta, \alpha_{\bar\eta} )  - i \int \wt{\boldsymbol\mu} \cdot \boldsymbol\alpha \right).
\label{effectiveaction}
\ee

\bigskip

In the rest of this paper we focus on calculations in the large N limit that 
are entirely determined by the saddle point values of the 
singlet fields.
We will be interested in saddle point solutions that preserve both 
translational invariance and have fermion number zero. 
In this case  the singlet fields become local\footnote{The corresponding formulas  for the bilocal fields are 
\beal{ \label{condm}
\aver{\chi(P,q)}=(2 \pi)^3 \delta^3(P) \chi(q), \;
\aver{\xi_{\bar\mu}(P,q)}= (2 \pi)^3 \delta^3(P) \xi_{\bar \delta}(q),\;
\aver{\bar{\eta} (P,q)}=\aver{ \eta (P,q)}= 0,
}
where $\bar \delta=I, -$. }
\beal{ \label{condm2}
&\aver{\alpha_B(P,q)}=(2 \pi)^3 \delta^3(P) \alpha_B(q), \;
\aver{\alpha_{F,\bar\delta}(P,q)}= (2 \pi)^3 \delta^3(P) \alpha_{F,\bar\delta}(q),\;
\aver{\alpha_{{\eta}} (P,q)}=\aver{ \alpha_{\bar{\eta}} (P,q)}= 0, \\
&\aver{i\mu_B(P,q)}=(2 \pi)^3 \delta^3(P) \Sigma_B(q), \;
\aver{i\mu_{F,\bar\nu}(P,q)}= (2 \pi)^3 \delta^3(P) \Sigma_{F,\bar\nu}(q),\;
\aver{i\mu_{{\eta}} (P,q)}= \aver{\Sigma_{\bar{\eta}} (P,q)}= 0, 
}
where $\bar \delta=I, -$ and $\bar \nu=I, +$. 
Using this it is simpler to derive the saddle point equations. The effective action (\ref{effectiveaction}) takes a simpler form, as we explain below.

 The first term in \eqref{effectiveaction} becomes 
\beal{ 
\STr\log Q(\boldsymbol\Sigma) 
&=\STr\log  \left(
\begin{array}{cc}
(q_s^2+q_3^2 +\Sigma_B(q) ) \delta(P)  & 0 \\
0 & ( i \gamma^\mu q_\mu + \gamma^+ \Sigma_{F,+}(q) +\Sigma_{F,I}(q)) \delta(P)
\end{array}\right) \nn
&=V \int \f{d^3q}{(2 \pi)^3} \left( \log (q_s^2+q_3^2 +\Sigma_B(q) )
-\log \det  ( i \gamma^\mu q_\mu + \gamma^+ \Sigma_{F,+}(q) +\Sigma_{F,I}(q)) \right),
}
where we set $V = (2 \pi)^3 \delta^3(0)$.
The second term in \eqref{effectiveaction} reduces to
\beal{
\wt V
&= V\biggl\{\int \f{d^3q_1}{(2 \pi)^3} \f{d^3q_2}{(2 \pi)^3}\f{d^3q_3}{(2 \pi)^3} C_2(q_1,q_2,q_3)
\alpha_B(q_1)\alpha_B(q_2)\alpha_B(q_3) \nn
&+\int  \f{d^3q_1}{(2 \pi)^3} \f{d^3q_2}{(2 \pi)^3}~
\f{8 \pi i \lambda}{  (q_1-q_2)_{-}} \alpha_{F,-}(q_1) \alpha_{F,I}(q_2) + \int \f{d^3q_1}{(2 \pi)^3} \f{d^3q_2}{(2 \pi)^3}
8\pi \lambda x_4 \alpha_B(q_1) \alpha_{F,I}(q_2) \biggl\},\label{tildeV}
}
where 
\be
C_2(q_1,q_2,q_3) =C_2(0,0,q_1,q_2,q_3)
={4 \pi^2 \lambda^2} \left(\f{(q_1+q_3)_{-} (q_2+q_3)_{-}}{(q_1-q_3)_{-} (q_2-q_3)_{-}} + x_6 \right).
\label{c2}
\ee
In order to obtain (\ref{tildeV}) we have used the fact that  the first term in \eqref{s1'} vanishes due to antisymmetry of the 
arguments inside the integral. Also note that the last term in \eqref{s1'},  $\Delta S_{BF}(\alpha_{\eta},\alpha_{\bar \eta})$ 
 vanishes 
by using (\ref{condm2}). 
The third term in \eqref{effectiveaction} which is written in (\ref{mualpha}) becomes 
\beal{
- i \int  \wt{\boldsymbol \mu} \cdot {\boldsymbol \alpha} 
&=  V\int  \f{d^3q}{(2 \pi)^3} 
\biggl(-\Sigma_B (q) \alpha_B(q) -2 \Sigma_{F,+}(q) \alpha_{F,-}(q)-2 \Sigma_{F,I}(q) \alpha_{F,I}(q)
\biggr).
}
Collecting all the terms we obtain 
\beal{
S_{eff} &= NV \biggl\{ \int \f{d^3q}{(2 \pi)^3} \left( \log (q_s^2+q_3^2 +\Sigma_B(q) )
-\log \det  ( i \gamma^\mu q_\mu + \Sigma_{F}(q)) \right) \nn
&+\int \f{d^3q_1}{(2 \pi)^3} \f{d^3q_2}{(2 \pi)^3}\f{d^3q_3}{(2 \pi)^3} C_2(q_1,q_2,q_3)
\alpha_B(q_1)\alpha_B(q_2)\alpha_B(q_3) \nn
&+\int  \f{d^3q_1}{(2 \pi)^3} \f{d^3q_2}{(2 \pi)^3}~
\f{8 \pi i \lambda}{  (q_1-q_2)_{-}} \alpha_{F,-}(q_1) \alpha_{F,I}(q_2) + \int \f{d^3q_1}{(2 \pi)^3} \f{d^3q_2}{(2 \pi)^3}
8\pi \lambda x_4  \alpha_B(q_1) \alpha_{F,I}(q_2) \nn
&+ \int  \f{d^3q}{(2 \pi)^3} 
\biggl(-\Sigma_B (q) \alpha_B(q) -2 \Sigma_{F,+}(q) \alpha_{F,-}(q)-2 \Sigma_{F,I}(q) \alpha_{F,I}(q)
\biggr) \biggr\},
\label{action-assumption}
}
where
\beal{
\Sigma_F &= \Sigma_{F,+} \gamma^+ +\Sigma_{F,-}\gamma^{-}+\Sigma_{F,3}\gamma^{3}+\Sigma_{F,I}I,\quad \mbox{with} \quad \Sigma_{F,-} = \Sigma_{F,3}=0.
}

As we noted before (see (\ref{alphapropa})), at the saddle point of \eqref{action-assumption} the singlet $\alpha$ and ${\Sigma}$ fields have a simple interpretation as the propagator and the self energy, 
respectively.
This follows immediately from the saddle point equations obtained 
by varying the action \eqref{action-assumption} with respect to the fields 
${\Sigma}_B$, ${\Sigma}_{F}$ respectively.%
\footnote{
The equation $2(\Sigma_{F,+} \alpha_{F,-}+\Sigma_{F,I} \alpha_{F,I})=\tr ({\Sigma_F} \alpha_F)$ will be useful for deriving the equation.
}
We obtain
\begin{equation}\label{pis}
 \alpha_{B}(q) = \frac{1}{q^{2}+{\Sigma}_{B}(q)}, \quad
\alpha_F(q)= - \frac{1}{i \gamma^\mu q_\mu +{\Sigma}_F(q)},
\end{equation}
where we have set 
\beal{
\alpha_F &= \alpha_{F,+} \gamma^+ +\alpha_{F,-}\gamma^{-}+\alpha_{F,3}\gamma^{3}+\alpha_{F,I}I,\quad \mbox{with} \quad\alpha_{F,+} = \alpha_{F,3}=0.
}

Using \eqref{pis} one can rewrite the remaining saddle point equations for $\alpha_B$, $\alpha_{F,I}$ and $\alpha_{F,-}$ 
 as integral equations for $\Sigma$ fields in the following way.
\begin{eqnarray}\label{gapeqnall}
{\Sigma}_{F,+}(p)&=&-{2 \pi i \lambda}\int \frac{d^3q}{(2\pi)^3}\frac{1}{(p-q)_{-} } \tr\frac{1}{i \gamma^\mu q_\mu +{\Sigma}_F(q)},\nonumber\\
{\Sigma}_{F,I}(p)&=&{2\pi i \lambda}\int \frac{d^3q}{(2\pi)^3} \frac{1}{(p-q)_{-}}
\tr\frac{\gamma^{-}}{i \gamma^\mu q_\mu +{\Sigma}_F(q)}
+4\pi \lambda x_{4} \int \frac{d^{3}q}{(2\pi)^3} \frac{1}{q^2+{\Sigma}_{B}(q)},\nonumber\\ 
{\Sigma}_{B}(p) &=&\int \frac{d^{3}q}{(2\pi)^3}\frac{d^{3}q'}{(2\pi)^3} \Big\lbrack C_2(p,q,q')+C_2(q,p,q')+ C_2(q,q',p)\Big\rbrack\f{1}{q^2+{\Sigma}_{B}(q)}\f{1}{q'^2+{\Sigma}_{B}(q')}\nonumber\\
&-&4\pi \lambda x_{4} \int \frac{d^3 q}{(2\pi)^3}\tr\frac{1}{i \gamma^\mu q_\mu +{\Sigma}_{F}(q) }.
\end{eqnarray}
We refer to these equations as   gap 
equations.

As a check on our algebra we will rederive the gap equations   (\ref{gapeqnall}) in two different ways. One is by using large $N$ factorization on Schwinger Dyson equations of the fundamental fields in Appendix \ref{SD} and the other by summing the planar diagrams contributing to 1PI self-energy diagrams, which will be studied in Appendix \ref{diagram}.

By using the equations of motion \eqref{pis} and \eqref{gapeqnall}
 we can rewrite (\ref{action-assumption})  in terms of $\Sigma_B, \Sigma_F$ as
\beal{\label{finansfe}
  S_{eff}&=NV\biggl[ \int \f{d^3q}{(2 \pi)^3} \biggl\{ \log (q_s^2+q_3^2 +\Sigma_B(q) )
-\log \det  ( i \gamma^\mu q_\mu + \Sigma_{F}(q)) \nn
&-\frac{2}{3} \frac{{\Sigma}_{B}(q)}{q_s^2+q_3^2 +{\Sigma}_{B}(q)}+ \frac{1}{2}\tr\left({\Sigma}_{F}(q)\frac{1}{i \gamma^\mu q_\mu + {\Sigma}_F(q)}\right) 
\biggr\}\nonumber\\
&- \frac{4\pi \lambda x_{4}}{6}
\left(\int\frac{d^{3}q}{(2\pi)^3}\frac{1}{q^{2}+{\Sigma}_{B}(q)}\right)
\left(\int\frac{d^{3}p}{(2\pi)^3} \tr\left(\frac{1}{i \gamma^\mu p_\mu + {\Sigma}_F(p) }\right)\right)
\biggr].
}

This is our answer of the exact effective action in the 't Hooft limit. 
This means the effective action is exact in all orders of $\lambda$. 
In a diagrammatic point of view, this effective action should encode the connected vacuum graph consisting of planar diagrams. 
We will check this point in Appendix \ref{diagram}.

\subsection{Exact solution of the gap equation}
\label{solution}

In this subsection we solve the gap equation \eqref{gapeqnall}. 
We note that 
the right-hand side of \eqref{gapeqnall} does not depend on $p_3$. 
This means that $\Sigma_B(p), \Sigma_F(p)$ are functions of
momenta $p_{+}$ and $p_{-}.$ We first concentrate on the bosonic case, which is given by the last equation in \eqref{gapeqnall}.
Due to rotational invariance the self energy for boson only depends on the combination $p_s=\sqrt{2p_+p_-}$. One can reach  the  
same conclusion by noting that the bosonic self-energy $\Sigma_B(p)$ is real. 
This  can be confirmed from the equation 
\be
\aver{\bar\phi(-q) \phi(p)} = (2 \pi)^3 N\delta^3(-q+p) \alpha(p)
={(2 \pi)^3 N\delta^3(-q+p) \over p^2 +\Sigma_B(p)},
\ee
which follows from \eqref{pis}.
Now\footnote{This method for solving the gap equation was first explained to us in the context of the purely fermionic theory 
 by S. Giombi. We thank him for discussions.} let us take the derivative with respect to $p_+$ on both sides of the last equation in (\ref{gapeqnall}). By using
\begin{equation}
 \f{\p}{\p p_{+}} \f{1}{p_{-}}= 2\pi \delta^2(p) ~~,~~\f{\p}{\p p_{+}} p_{-}=0 ~~\mbox{and}~~\f{\p}{\p p_{+}} p_{s}=\f{p_-}{p_s},
\label{delpplus} 
\end{equation} 
we find that the first term on the right-hand side vanishes. 
The second term proportional to $\lambda_4$  is independent of the external momentum
and therefore obviously  does not contribute. 
Therefore we obtain
\be\label{derisig}
\f{\p}{\p p_+}\Sigma_{B}(p)= 0. 
\ee
As a result $\Sigma_{B}(p)$ cannot depend on the momentum.
Since we do not have any other dimensionful parameter, 
we conclude that
\begin{equation}
 \Sigma_{B}(p)=0. \label{selfbos}
\end{equation} 

\bigskip

Now we turn to solving the gap equation for fermionic case. 
For this purpose, 
it is helpful to compare with  the purely fermionic case studied in \cite{Giombi:2011kc}.
 The difference between the purely fermionic case and our case is only the term with $\lambda_4$, 
which provides a constant shift independent of the external momentum in the equation for $\Sigma_{F,I}$. $\Sigma_{F,+}$ remains
the  same as that for the  purely fermionic case. 
The term proportional to $\lambda_4$ evaluates to zero by dimensional regularization
\begin{equation}
 \int \f{d^{3}q}{(2\pi)^3}\f{1}{q^2}=0,
\end{equation}where we have used $\Sigma_B =0.$
 Therefore the situation completely reduces to the  purely fermionic  case for which the 
 solution was already obtained in \cite{Giombi:2011kc} to be  
\begin{equation}
 \Sigma_{F} = f_{0}p_{s}I+ i g_{0}p_{+}\gamma^{+},~~ f_{0} = \lambda,~~g_{0} = -\lambda^2.
\label{solutionfermion}
\end{equation} 

We explicitly check that (\ref{selfbos}), (\ref{solutionfermion}) satisfy the equation of motion (\ref{gapeqnall}) 
in Appendix \ref{cgesol1}.

\section{Finite temperature}
\label{fta}

\subsection{Set up}
In this section we study $SU(N)$ level $k$ Chern-Simons theory with a fundamental matter at nonzero temperature, $T$. 
To this end we study the system on $\bR^2 \times \bS^1$ with circumference $\beta = 1/T$. 
The boundary conditions are  chosen to be periodic for the scalar and anti-periodic for the fermion.

The  important point is that the results for 
our computation on $\bR^3$ in the previous section  can be easily adapted to the  $\bR^2\times\bS^1$ case by  replacing the 
integration over the  3rd component of the momentum with  a  discrete sum determined by boundary condition. 
For terms which involved internal scalar propagators we do replacement 
\be
\int \frac{dp_3}{(2 \pi)} F(p_3) \to 
 \frac{1} {\beta} \sum_{p_3:B} F(p_3) := \frac{1} {\beta}
\sum_{n \in \bZ} F(\frac{2 \pi n}{\beta}), 
\ee
and for the part coming from an internal fermionic propagator, 
\be
\int \frac{dp_3}{(2 \pi)} F(p_3) \to 
 \frac{1} {\beta}
\sum_{p_3:F} F(p_3) := \frac{1} {\beta}
\sum_{n \in \bZ} F(\frac{2 \pi (n+\half)}{\beta}).
\ee
Note that the volume of space $V$ is now given by 
$V= V_2 \beta,$
where $V_2$ is the volume of 2-plane.

\subsection{Exact self energy}

In this subsection, we solve the gap equation at nonzero temperature. 
Using the replacement rules described above the gap equations 
become: 
\beal{\label{gapeqnfinitemp}
{\Sigma}_{F,+}(p)&=-{2 \pi i \lambda}\frac{1}{\beta}\sum_{q_3:F} \int \frac{d^2q}{(2\pi)^2}\frac{1}{(p-q)_{-} } \tr\frac{1}{i \gamma^\mu q_\mu +{\Sigma}_F(q)},\nonumber\\
{\Sigma}_{F,I}(p)&=-{2\pi i \lambda}\frac{1}{\beta}\sum_{q_3:F} \int \frac{d^2q}{(2\pi)^2} \frac{1}{(p-q)_{-}}
\tr\frac{\gamma^{-}}{i \gamma^\mu q_\mu +{\Sigma}_F(q)}
+4\pi\lambda x_{4} \frac{1}{\beta}\sum_{q_3:B} \int \frac{d^{2}q}{(2\pi)^2} \frac{1}{q^2+{\Sigma}_{B}(q)},\nonumber\\ 
{\Sigma}_{B}(p) &=\frac{1}{\beta^2}\sum_{q_3:B}  \int \frac{d^{2}q}{(2\pi)^2} 
\sum_{q'_3:B}\int \frac{d^{2}q'}{(2\pi)^2} \Big\lbrack C_2(p,q,q')+C_2(q,p,q')+ C_2(q,q',p)\Big\rbrack\f{1}{q^2+{\Sigma}_{B}(q)}\f{1}{q'^2+{\Sigma}_{B}(q')}\nonumber\\
&-4\pi \lambda x_{4} \frac{1}{\beta}\sum_{q_3:F}\int \frac{d^2 q}{(2\pi)^2}\tr\frac{1}{i \gamma^\mu q_\mu +{\Sigma}_{F}(q) }.
}

Our strategy for solving these equations will be similar to that used   in the zero temperature case in section \ref{solution}.
We start with the third equation in  (\ref{gapeqnfinitemp}) which determines $\Sigma_B$. 
The only dependence on the external momentum on the rhs of this equation is through the coefficients $C_2(p,q,q')$ which are 
independent of the third component of the momentum. An argument similar to that in the zero temperature case  shows that 
the dependence of the rhs on $p_+, p_-$ also vanishes. It therefore follows that $\Sigma_B$ must only depend on the temperature
and we can set 
\be
\Sigma_B(p) = \sigma T^2,
\label{ansatzsigmab}
\ee
where $\sigma$ is a  constant with a non-trivial dependence on $\lambda$ which needs to be determined.

Next we turn to the first two equations in  (\ref{gapeqnfinitemp}) which determine $\Sigma_{F,I}, \Sigma_{F,+}$.
Once again the rhs of these two equations do not depend on $p_3$. Rotational invariance in the 1-2 plane and 
dimensional analysis then allow us to write  
\be
\Sigma_{F,I}(p) = f(\wt p) p_s, \quad 
\Sigma_{F,+}(p) =i g(\wt p) p_+,
\label{ansatzsigmaf}
\ee
where $f(\wt p), g(\wt p)$ are undetermined functions of $\wt p={p_s \over T}$, which is dimensionless.


\subsubsection{Self energy of fermionic field}

Using the ansatz in  (\ref{ansatzsigmaf})  the first two equations in \eqref{gapeqnfinitemp} take the form
\beal{
g(\wt p) p_+&=-{4 \pi  \lambda}\frac{1}{\beta}\sum_{q_3:F} \int \frac{d^2q}{(2\pi)^2}
\frac{1}{(p-q)_{-} } \frac{f(\wt q) q_s}{ (1+g(\wt q)+f(\wt q)^2)q_s^2 +q_3^2 }, \label{intg1}\\
f(\wt p) p_s&={4\pi \lambda}\frac{1}{\beta}\sum_{q_3:F} \int \frac{d^2q}{(2\pi)^2}\frac{1}{(p-q)_{-} } \frac{q_-}{ (1+g(\wt q)+f(\wt q)^2)q_s^2 +q_3^2 }
+4\pi \lambda x_{4} \frac{1}{\beta}\sum_{q_3:B} \int \frac{d^{2}q}{(2\pi)^2} \frac{1}{q^2+\sigma T^2} \label{intf1}. 
}
Now let us take the derivative with respect to $p_+$ on both sides. By using \eqref{delpplus} we find 
\beal{
{\partial \over\partial p_+} (g(\wt p) p_+)&=-{4 \pi \lambda}\frac{1}{\beta }\sum_{q_3:F} 
{1\over (2\pi)} \frac{f(\wt p) p_s}{ (1+g(\wt p)+f(\wt p)^2)p_s^2 +q_3^2 },\\
{\partial \over\partial p_+} (f(\wt p) p_s)&={4\pi \lambda}\frac{1}{\beta}\sum_{q_3:F} 
{1\over (2\pi)}  \frac{p_-}{ (1+g(\wt p)+f(\wt p)^2)p_s^2 +q_3^2 }. 
}
Combining these equations we find
\be
{\partial \over\partial p_+} \left( (f(\wt p) p_s)^2 + g(\wt p) p_s^2 \right) =0,
\ee
which implies
\be
f(\wt p)^2 + g(\wt p) = {c \over \wt p^2},
\label{fgc}
\ee
where $c$ is a dimensionless constant with a nontrivial dependence on $\lambda$. 
From this relation, $g$ can be obtained once we determine $f$. 
Therefore, we focus on solving the equation for $f$. 
Plugging \eqref{fgc} back, we find 
\beal{
f(\wt p) p_s&={4\pi \lambda}\frac{1}{\beta}\sum_{q_3:F} \int \frac{d^2q}{(2\pi)^2}\frac{1}{(p-q)_{-} } \frac{q_-}{ q_s^2 +q_3^2 +cT^2 }
+4\pi \lambda x_{4} \frac{1}{\beta}\sum_{q_3:B} \int \frac{d^{2}q}{(2\pi)^2} \frac{1}{q^2+\sigma T^2}. 
\label{f}
}
Let us first calculate the first term. 
First we do the angular part of the integration by using 
\be
 \int_0^{2\pi} \frac{d\phi}{2\pi} \frac{q_-}{(p-q)_{-} } = - { \theta(q_s-p_s)},
\ee 
where $\phi$ is the angular part of $q_-$, $q_- =q_s e^{i \phi}$, and $\theta(x)$ is a step function, $\theta(x)=1$ for $x>0$, otherwise vanishes. 
Then we obtain
\beal{
\mbox{(1st term in \eqref{f})} &={4\pi \lambda}\frac{1}{\beta}\sum_{q_3:F} 
\int_{p_s}^\infty \frac{q_s dq_s}{(2\pi)} \frac{-1}{ q_s^2 +q_3^2 +cT^2 }. 
}
Since this integral includes UV divergence, we regularize this integral as explained in Appendix \ref{regularization}. 
After the regularization, we find 
\beal{
\mbox{(1st term in \eqref{f})} &={4\pi \lambda}\times { 1 \over 2\pi \beta} \log(2 \cosh (\frac{\sqrt{\wt p^2 +c}}{2}) ). 
}
Detailed calculation is available in Appendix \ref{formulas}.

The second term in \eqref{f} can also be computed in a similar manner. 
 Details of the calculation can be found in to Appendix \ref{formulas} with the result 
\beal{
\mbox{(2nd term in \eqref{f})} &=4\pi \lambda x_4 \times { -1 \over 2\pi \beta} \log(2 \sinh (\frac{\sqrt{\sigma}}{2}) ).
}
The function $\sqrt{\sigma}$ which appears in this formula is to be taken as the positive root. 
 Also in obtaining this formula we 
are assuming  $\sigma\ge 0$. We will see below that this assumption is met by our solution in the ${\cal N}=2$ case
 for all values of $\lambda$. 
 Summing up these contributions, we find the solution of $f$ and thus $g$ as follows. 
\bea
f (\wt p) &=& \frac{2\lambda}{\wt p}
\biggl(\log(2\cosh(\frac{\sqrt{\wt p^2 + c}}{2})) - x_4 \log({2\sinh(\frac{\sqrt\sigma}{2})})\biggr). \label{fgsola} \\
g(\wt p) &=&{c \over \wt p^2}- f(\wt p)^2. 
\label{fgsol}
\eea
Here, again for convenience, we set $\wt p = {\beta p_s}$. 

\paragraph{On determining the constant $c$ in terms of $\sigma$ }
The solution for the self energy obtained above has two  constants $c$ and $\sigma$. 
We now argue that the behavior of the integral equations (\ref{intg1}), (\ref{intf1}) 
 give rise to one relation between them. 
For this purpose, let us study the behavior of $g(\wt p)$ around $p\sim0$. 
From (\ref{fgsola}), (\ref{fgsol}), we find 
\be
g(\wt p) =  {c - \left\{ 2\lambda \left( \log(2\cosh(\frac{\sqrt{ c}}{2})) - x_4 \log({2\sinh(\frac{\sqrt\sigma}{2})}) \right) \right\}^2 \over \wt p^2} + \cdots. 
\label{garound0}
\ee
On the other hand \eqref{intg1}, \eqref{fgsola} imply that 
$g(\wt p) \sim \cO(1)$ around $\wt p\sim0$. 
To see this let us evaluate \eqref{intg1} by performing the summation over $q_3$ and the integral of the angular part of $q$. 
\beal{
g(\wt p) p_+&=-{4 \pi  \lambda}  \int \frac{d^2q}{(2\pi)^2}
\frac{1}{(p-q)_{-} } \frac{\tanh( {\beta\sqrt{q_s^2 +cT^2}\over 2}) }{ 2\sqrt{q_s^2 +cT^2}} f(\wt q) q_s\\
&=-{4 \pi  \lambda}  \int_0^{p_s} \frac{q_s dq_s}{2\pi}
\frac{1}{p_{-} } \frac{\tanh( {\beta\sqrt{q_s^2 +cT^2}\over 2}) }{ 2\sqrt{q_s^2 +cT^2}}f(\wt q) q_s.
}
Here we used \eqref{fgc}, \eqref{sumform2} and 
\be
 \int_0^{2\pi} \frac{d\phi}{2\pi} \frac{1}{(p-q)_{-} } = { \theta(p_s-q_s) \over p_-}.
\ee 
Therefore we obtain
\beal{
g(\wt p) 
&=-{4 \pi  \lambda \over p_s^2}  \int_0^{p_s} \frac{q_s dq_s}{2\pi}
\frac{\tanh( {\beta\sqrt{q_s^2 +cT^2}\over 2}) }{ \sqrt{q_s^2 +cT^2}}f(\wt q) q_s.
}
Now \eqref{fgsola} implies that $f (\wt p) \sim {1 \over \wt p}$ around $p \sim 0$, 
we find that $g(\wt p) $ behaves as $g(\wt p)  \sim \cO(1)$ around $p \sim 0$. 
Therefore, the singular term in terms of momentum in \eqref{garound0} has to vanish and we obtain
\be
c = \left\{ 2\lambda \left( \log(2\cosh(\frac{\sqrt{ c}}{2})) - x_4 \log({2\sinh(\frac{\sqrt\sigma}{2})}) \right) \right\}^2.
\label{girfree}
\ee


\subsubsection{Self energy of scalar field}

Next we evaluate the self energy for the scalar field using  the third equation of \eqref{gapeqnfinitemp}. 
Since this equation does not depend on the external momentum $p$, we can set $p$ to zero. 
\beal{
\sigma T^2 =&\frac{1}{\beta^2}\sum_{q_3:B}  \int \frac{d^{2}q}{(2\pi)^2} 
\sum_{q'_3:B}\int \frac{d^{2}q'}{(2\pi)^2} \Big\lbrack C_2(0,q,q')+C_2(q,0,q')+ C_2(q,q',0)\Big\rbrack\nn
&\times\f{1}{q^2+\sigma T^2}\f{1}{q'^2+\sigma T^2}
-4\pi\lambda x_{4} \frac{1}{\beta}\sum_{q_3:F}\int \frac{d^2 q}{(2\pi)^2}\tr\frac{1}{i \gamma^\mu q_\mu +{\Sigma}_{F}(q) }.
}
For convenience we denote the first term by $\Sigma_1$ and the second term by $\Sigma_2$. 
First we focus on $\Sigma_1$, which takes the form, \eqref{c2},
\beal{\label{sigma1}
\Sigma_1&= 4 \pi^2 \la^2(1+ 3 x_6) \f{1}{\beta^2} \sum_{n,m} \int \f{d^2q}{(2\pi)^2}\f{d^2q'}{(2\pi)^2} \f{1}{\lbrack(\f{2 \pi n}{\beta})^2+q_s^2+\sigma T^2\rbrack}\f{1}{\lbrack(\f{2 \pi m}{\beta})^2+q_s^{'2}+\sigma T^2\rbrack} \nonumber \\
&- 8 \pi^2 \la^2 \f{1}{\beta^2}\sum_{n,m}\int \f{d^2q}{(2\pi)^2}\f{d^2q'}{(2\pi)^2} \f{(q+q')_{-}}{ (q-q')_{-}} \f{1}{\lbrack(\f{2 \pi n}{\beta})^2+q_s^2+\Sigma_{B}(q)\rbrack}\f{1}{\lbrack(\f{2 \pi m}{\beta})^2+q_s^{'2}+\sigma T^2\rbrack}.
}
Notice that the second term in $\Sigma_1$ vanishes due to the fact that the integrand is odd with respect to $q, q'$. 
Therefore, we need only to evaluate the first term, 
which can be obtained by applying the formula \eqref{form2} twice, derived in Appendix \ref{formulas}.  
Thus we find 
\beal{
\Sigma_1&= 4 \pi^2 \la^2(1+ 3 x_6) \left(-{ 1 \over 2\pi \beta} \log(2 \sinh (\frac{\sqrt{\sigma}}{2}) )\right)^2.
}
Next we evaluate $\Sigma_2$. 
\be
\Sigma_2= -4\pi\lambda x_{4} \frac{1}{\beta}\sum_{n}\int \frac{d^2 q}{(2\pi)^2} 
 \f{2\Sigma_{F,I}(q)}{(\f{2 \pi (n+\half)}{\beta})^2+q_s^2+cT^2}.
\label{sigma2}
\ee
Substituting $\Sigma_{F,I}(q) = f(\wt q) q_s$ into this we find 
\be
\Sigma_2= -4\pi\lambda x_{4}  \frac{4\lambda}{\beta} \left(I_1 -x_4 \log(2 \sinh({\sqrt{\sigma} \over 2})) I_0 \right),
\ee
where we set
\beal{
I_1 &= \frac{1}{\beta} \sum_{n}\int \frac{d^2 q}{(2\pi)^2} 
 \f{\log(2\cosh({\sqrt{(\beta q_s)^2+c}\over 2})) }{(\f{2 \pi (n+\half)}{\beta})^2+q_s^2+cT^2}, \quad
I_0 = \frac{1}{\beta} \sum_{n}\int \frac{d^2 q}{(2\pi)^2} 
 \f{1 }{(\f{2 \pi (n+\half)}{\beta})^2+q_s^2+cT^2}.
}
These integrals include divergence, so we have to regularize them to obtain finite results. We perform detailed calculation in
 Appendix \ref{formulas} (see \eqref{form3}, \eqref{form2}).  
The result is 
\beal{
I_1 &= -{ \left(\log(2\cosh({\sqrt c \over 2})) \right)^2    \over 4\pi \beta}, \quad
I_0 =-{ 1 \over 2\pi \beta} \log(2 \cosh (\frac{\sqrt{c}}{2}) ).
}
Therefore we find
\be
\Sigma_2= -  \frac{4\lambda^2 x_{4} }{\beta^2} \left(-{ \left(\log(2\cosh({\sqrt c \over 2})) \right)^2  } + 2 x_4 \log(2 \sinh({\sqrt{\sigma} \over 2})) \log(2 \cosh (\frac{\sqrt{c}}{2}) ) \right).
\label{sigma2answer}
\ee
Collecting these contributions, we find the solution for $\sigma$ as
\beal{
\sigma&= \lambda^2 \biggl[ (1+3 x_6)  \( \log(2 \sinh({\sqrt{\sigma} \over 2})) \)^{2}
-8 x_4^2  \log(2 \sinh({\sqrt{\sigma} \over 2})) \log(2 \cosh (\frac{\sqrt{c}}{2}) ) \nn
&\qquad + 4 x_4 \left(\log(2\cosh({\sqrt c \over 2})) \right)^2 \biggr].
\label{sigmasol}
}
On the rhs the function $\sqrt{\sigma}$ is to be taken as the positive root. 

Let us conclude this subsection by summarising the results obtained so far. 
The self energy for the boson is given in  (\ref{ansatzsigmab}) and for the fermion is given in  (\ref{ansatzsigmaf})
 with the functions $f,g$ being given in  (\ref{fgsola})  and  (\ref{fgsol}). These answers depend on two coefficients
$c, \sigma$ which are obtained by solving the two equations  (\ref{girfree}),  (\ref{sigmasol}). 
These expressions for the self energies are some of the main results  of this paper. 
 
\subsection{Exact free energy}

In this subsection we compute the free energy of the Chern-Simons-matter system on $\bR^2\times \bS^1$. 
For this purpose, we divide the free energy density $F$ into each contribution from scalar field, fermionic field and their interaction, denoted by $F_B, F_F, F_{BF}$, respectively. 
They are given by 
\be
F_B ={ \wt S_B(T) - \wt S_B(0) \over V}, \quad
F_F ={ \wt S_F(T) - \wt S_F(0) \over V}, \quad
F_{BF} ={ \wt S_{BF}(T) - \wt S_{BF}(0) \over V}, \quad
\ee
where $\wt S_B(T), \wt S_F(T), \wt S_{BF}(T)$ are respectively the contributions to the effective action coming from scalar field, fermionic field and their interaction on $\bR^2\times \bS^1$
\beal{
\wt S_B(T)&= NV  \frac{1} {\beta} \sum_{q_3:B} \int \f{d^2q}{(2 \pi)^2} \biggl(  \log\left(  {q^2 + \Sigma_B(q) } \right) - \frac{2}{3} \frac{{\Sigma}_{B}(q)}{q^{2}+{\Sigma}_{B}(q)} \biggl), \\
\wt S_F(T)&=  NV  \frac{1} {\beta} \sum_{q_3:F} \int \frac{d^2 q}{(2 \pi)^2} {\rm tr} \biggl( - \ln\left( {i \gamma^\mu q_\mu + \Sigma_F(q) } \right)+\frac{1}{2} {\Sigma}_{F}(q)\frac{1}{i \gamma^\mu q_\mu + {\Sigma}_F(q)} \biggl), \\
\wt S_{BF}(T)&=-\frac{4\pi \lambda x_{4}NV}{6}
 \frac{1} {\beta} \sum_{q_3:B}\int\frac{d^{2}q}{(2\pi)^2}\frac{1}{q^{2}+{\Sigma}_{B}(q)}
 \frac{1} {\beta} \sum_{p_3:F}\int\frac{d^{2}p}{(2\pi)^2}\tr\left(\frac{1}{i \gamma^\mu p_\mu + {\Sigma}_F(p) }\right),
}
and $\wt S_B(0), \wt S_F(0), \wt S_{BF}(0)$ are on $\bR^3$. 

In what follows we compute $F_B, F_F, F_{BF}$ separately.

\subsubsection{Contribution from scalar field}
First we compute $F_B$, which is given by 
\beal{ \label{fren}
 F_B =&N \biggl[ {1 \over \beta} \sum_{n} \int \frac{d^2 q}{(2 \pi)^2}
 \left( \ln\left( ({2\pi n \over \beta})^2 + q^2 + \sigma T^2 \right)
- \frac{2}{3} \frac{ \sigma T^2 }{ ({2\pi n \over \beta})^2 + q^2 + \sigma T^2} \right) \nn
&  - \int \frac{d^3 q}{(2 \pi)^3} \ln\left( q^2  \right)
\biggr].
}
Here we already used the solution $\Sigma_B=\sigma T^2$ on $\bR^2\times \bS^1$ and $\Sigma_B =0$ on $\bR^3$. 
We compute this in the following way. 
\bea \label{fren1}
&& F_B=N \biggl[ {1 \over \beta} \sum_{n} \int \frac{d^2 q}{(2 \pi)^2}
\ln\left(({2\pi n \over \beta})^2  + q^2 + \sigma T^2  \right)
- \int \frac{d^3 q}{(2 \pi)^3} 
\ln\left(  q^2  +\sigma T^2   \right) \notag\\
&&\qquad + \int \frac{d^3 q}{(2 \pi)^3} 
\log\big(\frac{ q^2  + \sigma T^2  }{ q^2}) - \frac{1}{ \beta} \sum_n \int \frac{d^2 q}{(2 \pi)^2} 
\frac{2}{3} \left( \frac{ \sigma T^2 }{({2\pi n \over \beta})^2 + q^2 +\sigma T^2 } \right) 
\biggr].
\eea

Since the first line in the bracket is the same as Casimir energy of free complex scalar fields with mass squared $\sigma T^2$, it can be calculated as 
\beal{
2 \times \int \frac{d^2 q}{(2 \pi)^2} ~ \log \left ( 1- e^{-(q^2+\sigma T^2)} \right) =&\frac{ T^3 }{ \pi} \int_{ \sqrt\sigma  }^\infty dy ~y \log \left ( 1- e^{-y} \right).
}

As first term in the second line of (\ref{fren1}) is linearly divergent, 
one can regularize it as explained in Appendix \ref{regularization} 
\beal{
\int \frac{d^3 q}{(2 \pi)^3} 
\log\big(\frac{ q^2  + \sigma T^2  }{ q^2}) \to 
\lim_{m\to0}\int \frac{d^D \hat q}{(2 \pi)^D} 
\log\big(\frac{\hat q^2  + \sigma T^2  }{\hat q^2 + m^2}).
}
We evaluate this as follows. 
\beal{
\int \frac{d^D \hat q}{(2 \pi)^D} 
\log\big(\frac{\hat q^2  + \sigma T^2  }{\hat q^2 + m^2})
&= \int \frac{d^{D-1} \hat q }{(2 \pi)^{D-1}} ( \sqrt {\hat q^2  + \sigma T^2} -  \sqrt {\hat q^2  + m^2})  \nn
&= {1 \over (4\pi)^{D-1 \over 2}} {\Gamma(-\half - {D-1 \over 2}) \over \Gamma(-\half)} \left( (\sigma T^2)^{D \over 2} - (m^2)^{D \over 2}   \right) \nn
&= -{1 \over  6 \pi} ( T^3 \sigma^{3 \over 2} -m^3)  +\cO(\epsilon), 
}
where $D=3-\epsilon$ is negative to make this integral convergent. 
Analytically continuing from some positive number $\epsilon$ to zero, 
we evaluate this as $-{1 \over  6 \pi} ( T^3 \sigma^{3 \over 2} -m^3)$, which 
becomes $-{1 \over  6 \pi} ( T^3 \sigma^{3 \over 2})$ under $m\to0$.

By using the formula \eqref{form1p0} one can evaluate the second term in the second line of (\ref{fren1}),   which gives 
\begin{equation}
\frac{T^3}{3 \pi}  \sigma \ln \left(2 \sinh\left( { \sqrt \sigma  \over 2}\right)\right).
\end{equation}

Collecting these terms we obtain 
\be
 F_B = 
 \frac{N T^3}{6 \pi} \biggl( - \sigma ^{3\over 2} 
+ 2  \sigma \log (2 \sinh( { \sqrt \sigma  \over 2}))
 + 6\int_{\sqrt \sigma}^\infty dy ~ y \ln \left ( 1- e^{-y} \right) 
 \biggr).\label{fb}
\ee

Note that the expression \eqref{fb} is the same as the free energy density of Chern-Simons-scalar theory.

\subsubsection{Contribution from fermionic field}

Next we concentrate on the contribution to free energy density coming from the fermions, $F_F$.
\beal{
F_F &=N \frac{1} {\beta} \sum_{q_3:F} \int \frac{d^2 q}{(2 \pi)^2} {\rm tr} \biggl( - \ln\left( {i \gamma^\mu q_\mu + \Sigma_F(q) } \right)+\frac{1}{2} {\Sigma}_{F}(q)\frac{1}{i \gamma^\mu q_\mu + {\Sigma}_F(q)} \biggl) \nn
&-N \int \frac{d^3 q}{(2 \pi)^3} {\rm tr} \biggl( - \ln\left( {i \gamma^\mu q_\mu + \Sigma^{(0)}_F(q) } \right)+\frac{1}{2} {\Sigma}_{F}^{(0)}(q)\frac{1}{i \gamma^\mu q_\mu + {\Sigma}^{(0)}_F(q)} \biggl),
}
where $\Sigma_F, \Sigma_F^{(0)}$ are the self energy of fermions on $\bR^2\times \bS^1$, $\bR^3$, respectively. 
We compute this as follows. 
\beal{
F_F&=N\biggl[ \frac{1} {\beta} \sum_{n} \int \frac{d^2 q}{(2 \pi)^2} 
 - \log\left( ({2\pi (n+\half) \over \beta})^2 + q_s^2 + cT^2 \right)
+ \int \frac{d^3 q}{(2 \pi)^3}  \log \left(q_3^2 + q_s^2 + cT^2\right) \nn 
&\qquad- \int \frac{d^3 q}{(2 \pi)^3}  \log \left({ q_3^2 + q_s^2 + cT^2 \over q_3^2 +q_s^2}\right)
+ \frac{1} {\beta} \sum_{n} \int \frac{d^2 q}{(2 \pi)^2} \frac{1}{2} \frac{\Sigma_{F,I}(q)^2 + cT^2}{ ({2\pi (n+\half) \over \beta})^2 + q_s^2 + cT^2 }  \nn
&\qquad - \int \frac{d^3 q}{(2 \pi)^3} \frac{1}{2}\frac{\Sigma^{(0)}_{F,I}(q)^2}{q_3^2 +q_s^2  } \biggl]. 
\label{Freeenergyfer}
}

The first-line in the bracket describes Casimir energy of free spin half fermions with mass squared $c T^2$, and thus it can be evaluated as 
\beal{
-2\times \int \frac{d^2 q}{(2 \pi)^2} ~ \log \left ( 1 + e^{-(q^2+c T^2)} \right) 
=&-\frac{ T^3 }{ \pi} \int_{ \sqrt c  }^\infty dy ~y \log \left ( 1 + e^{-y} \right).
}

The first term in the second line of (\ref{Freeenergyfer}) was already evaluated in the scalar case. 
The result is $\frac{T^{3} {c}^{3 \over 2}}{6\pi}.$

By substituting the solution of $\Sigma_F$, 
the second term in the second-line of (\ref{Freeenergyfer}) can be written as
\beal{
\half \left({2\lambda \over \beta}\right)^2 
\left[ I_2 
-2x_4 \log (2 \sinh( { \sqrt \sigma  \over 2}))I_1 
+ \left\{ \left( x_4 \log (2 \sinh( { \sqrt \sigma  \over 2}))\right)^2+cT^2 \right\}I_0
\right],
}
where $I_a\, (a=0,1,2)$ are given by 
\be
I_a=\frac{1}{\beta} \sum_{n}\int \frac{d^2 q}{(2\pi)^2} 
 \f{\left(\log\left(2\cosh({\beta\sqrt{q_s^2+cT ^2}\over 2})\right)\right)^a }{(\f{2 \pi (n+\half)}{\beta})^2+q_s^2+cT^2}.
\ee
These are calculated in Appendix \ref{formulas}, \eqref{form3}. The result is
\be
I_a
= -{ \left(\log\left(2\cosh({\sqrt c \over 2})\right) \right)^{a+1}    \over 2\pi \beta(a+1)}.
\ee

The final term in (\ref{Freeenergyfer}) is given by
\be
-{\lambda^2 \over 2} \int \frac{d^3 q}{(2 \pi)^3} \frac{q_s^2}{q_3^2 +q_s^2  },
\ee  
which can be computed in a similar manner to \eqref{j4}. 
One finds that this term vanishes as \eqref{j4} does.

Collecting these contributions we obtain
\beal{
F_F
=& {N T^3 \over 6 \pi}  \biggl[{ \sqrt c^3 }  - 6 \int_{\sqrt c}^\infty dy y \log (1 + e^{- y} ) - 2{\lambda^2 } \left(\log(2\cosh({\sqrt{c}\over 2}))\right)^3 \nn
&- {3  \over 2} c \log(2\cosh({\sqrt{c}\over 2})) 
+6 {x_4 \lambda^2 \log (2 \sinh( { \sqrt \sigma  \over 2}))\left(\log(2\cosh({\sqrt{c}\over 2}))\right)^2} \nn
&-6 { \lambda^2 x_4^2 \left(\log (2 \sinh( { \sqrt \sigma  \over 2}))\right)^2 \log(2\cosh({\sqrt{c}\over 2}))}  \biggr].
}

Note that when $x_4=0$ $F_F$ reduces to the free energy density of the purely fermionic system studied in \cite{Giombi:2011kc}.

\subsubsection{Contribution from interaction piece}

Finally we compute the contribution coming from the interaction between scalars and fermions. 
\beal{
F_{BF} &=-\frac{4\pi \lambda x_{4}N}{6}
\biggl( \frac{1} {\beta} \sum_{q_3:B}\int\frac{d^{2}q}{(2\pi)^2}\frac{1}{q_3^{2}+q_s^2+\sigma T^2}
 \frac{1} {\beta} \sum_{p_3:F}\int\frac{d^{2}p}{(2\pi)^2}\tr\left(\frac{1}{i \gamma^\mu p_\mu + {\Sigma}_F(p) }\right)  \nn
&\qquad- \int\frac{d^{3}q}{(2\pi)^3}\frac{1}{q_3^{2}+q_s^2}
\int\frac{d^{3}p}{(2\pi)^3}\tr\left(\frac{1}{i \gamma^\mu p_\mu + {\Sigma}_F^{(0)}(p) }\right) \biggr).\label{Freeenergyinter} 
}
We first compute the first-line. By using the formula \eqref{form1p0}, we obtain
\beal{ 
&-{4\pi \lambda x_{4} N \over 6} \left( - { 1 \over 2\pi \beta} \log(2 \sinh ( \frac{\sqrt{\sigma}}{2}) )\right) 
\times  \frac{1} {\beta} \sum_{n}\int\frac{d^{2}p}{(2\pi)^2} \frac{2 \Sigma_{F,I}(p) }{({2\pi(n+\half) \over \beta})^2 +p_s^2 +cT^2 } \nn
=&{N \over 6} \left( - { 1 \over 2\pi \beta} \log(2 \sinh ( \frac{\sqrt{\sigma}}{2}) )\right)\times \Sigma_2,  \label{Freeenergyinter1} 
}
where we used \eqref{sigma2} to obtain the second line. $\Sigma_2$ can be evaluated in a similar way  as in \eqref{sigma2answer},
 so we get the first-line as 
\beal{
-& { N \over 12\pi \beta} \log(2 \sinh ( \frac{\sqrt{\sigma}}{2}) ) \nn
&\times  \frac{4\lambda^2 x_{4} }{\beta^2} \left({ \left(\log(2\cosh({\sqrt c \over 2})) \right)^2  } - 2 x_4 \log(2 \sinh({\sqrt{\sigma} \over 2})) \log(2 \cosh (\frac{\sqrt{c}}{2}) ) \right).
}
We note that the second line in (\ref{Freeenergyinter}) vanishes (see  (\ref{j1},\ref{j2}) for details).
 Thus we obtain
\beal{
F_{BF}
=& -  {NT^3  x_4 \lambda^2 \over 3 \pi}  \biggl(  \log(2 \sinh({\sqrt{\sigma} \over 2}))  \left(\log(2\cosh({\sqrt c \over 2})) \right)^2 \nn
&\qquad\qquad\qquad  -2 x_4 \left( \log(2 \sinh({\sqrt{\sigma} \over 2})) \right)^2 \log(2\cosh({\sqrt c \over 2}))   \biggr).
}

\bigskip

Summing up all terms and simplifying we obtain the free energy density as follows. 
\beal{
F
&={NT^3 \over 6\pi}   \biggl[ -\sqrt\sigma^3 + { \sqrt c^3} +  2 \sigma  \log(2 \sinh({\sqrt{\sigma} \over 2}))  -{3\over2} c \log(2\cosh({\sqrt c \over 2})) \nn
&\qquad  -2 \lambda^2 \log(2\cosh({\sqrt c \over 2})) \left\{\log(2\cosh({\sqrt c \over 2})) -x_4  \log(2 \sinh({\sqrt{\sigma} \over 2})) \right\}^2 \nn
&\qquad +6 \int_{\sqrt \sigma}^\infty dy y \log ( 1-e^{-y}) 
-6 \int_{\sqrt c}^\infty dy y \log ( 1+ e^{-y} ) \biggr]. \label{finalfree} 
}
The functions $\sqrt{c}$ and $\sqrt{\sigma}$ which appear on the rhs stand for the positive values of the square root function. 

Th expression for the free energy density in  (\ref{finalfree}), which is valid to all orders in  the 't Hooft  coupling $\lambda$ and  the
other couplings,  $x_4$ and $x_6$, is one of the main results of this paper. 
Note that the coefficients $c, \sigma $ which appear in  (\ref{finalfree}) can be obtained in terms of the couplings and the 
temperature from  (\ref{girfree}) and  (\ref{sigmasol}).

\subsection{Supersymmetric case}
\label{susyapply}
\subsubsection{${\cal N}=2$ case}
So far  our analysis  has been carried out for general values of the couplings $x_4, x_6$ defined in \eqref{coefficients}.
One of our main motivations is to study the ${\cal N}=2$ SUSY theory in our setup and we turn to applying our general results to this theory
 now. For the ${\cal N}=2$ theory
$x_4,x_6$ satisfy the relation%
\footnote{The ${\cal N}=2$ theory has two additional couplings 
$x_4', x_4''$ defined in  (\ref{coefficients}) which  take values $x_4'=1, x_4''=0$ in the large $N$ limit. Their effects are suppressed in the large 
$N$ limit, however.}
\be
\label{relsusy}
x_4= x_6 = 1
\ee
in the large $N$ limit.
More details on the ${\cal N}=2$ theory can be found in appendix \ref{susycsn2}.

The self energy of the boson and the fermion was given in \eqref{ansatzsigmab}, \eqref{ansatzsigmaf}, \eqref{intg1}, \eqref{intf1},
 in terms of the constants $\sigma$ and $c$ which are determined by 
\eqref{sigmasol}, \eqref{girfree}. 
The equation \eqref{sigmasol}, which determines $\sigma$, for the choice of couplings \eqref{relsusy} reduces to
\be
\sigma = 4\lambda^2 \left( \log(2 \sinh({\sqrt{\sigma} \over 2})) -  \log(2 \cosh({\sqrt{c} \over 2}))\right)^2.
\ee
On the other hand \eqref{girfree} becomes
\be
c = 4\lambda^2 \left( \log(2 \cosh({\sqrt{c} \over 2}))-  \log(2 \sinh({\sqrt{\sigma} \over 2}))\right)^2.
\ee
From these we find\footnote{$\sqrt{\sigma}$, $\sqrt{c}$ in the solution stand for the positive root.}
\be
\label{scsusy}
\sqrt\sigma=\sqrt c= 2 |\lambda|  \log( \coth\frac{\sqrt{c}}{2}).
\ee
This implies that the thermal masses of the scalar and fermion become the same in $\cN=2$ supersymmetric case.
Correspondingly  there is also a simplification in the free energy density which in this case becomes%
\footnote{Note that we obtained the same results for $c$ \eqref{scsusy} and the free energy density \eqref{fesusy1} by momentum cut-off regularization. 
We checked that the divergences appearing in the gap equations \eqref{scsusy} and the free energy \eqref{fesusy1} in this regularization scheme cancel. See also \cite{Avdeev:1992jt} for cancellation of divergence in $\cN=2$ case. }
\beal{
F&= -{NT^3 \over 6\pi}  \biggl({ \sqrt c^3 \over |\lambda| } +
6 \int _{\sqrt c}^\infty dy y \log ( \coth {y \over 2} ) \biggr) \label{fesusy1} \\
&= -{NT^3 \over 6\pi}  \biggl( {21 \over 2} \zeta(3) + { \sqrt c^3 \over |\lambda| } -
6 \int^{\sqrt c} _0 dy y \log ( \coth {y \over 2} ) \biggr).
 \label{fesusy2}
}

\begin{figure}
  \begin{center}
  \subfigure[]{\includegraphics[scale=.5]{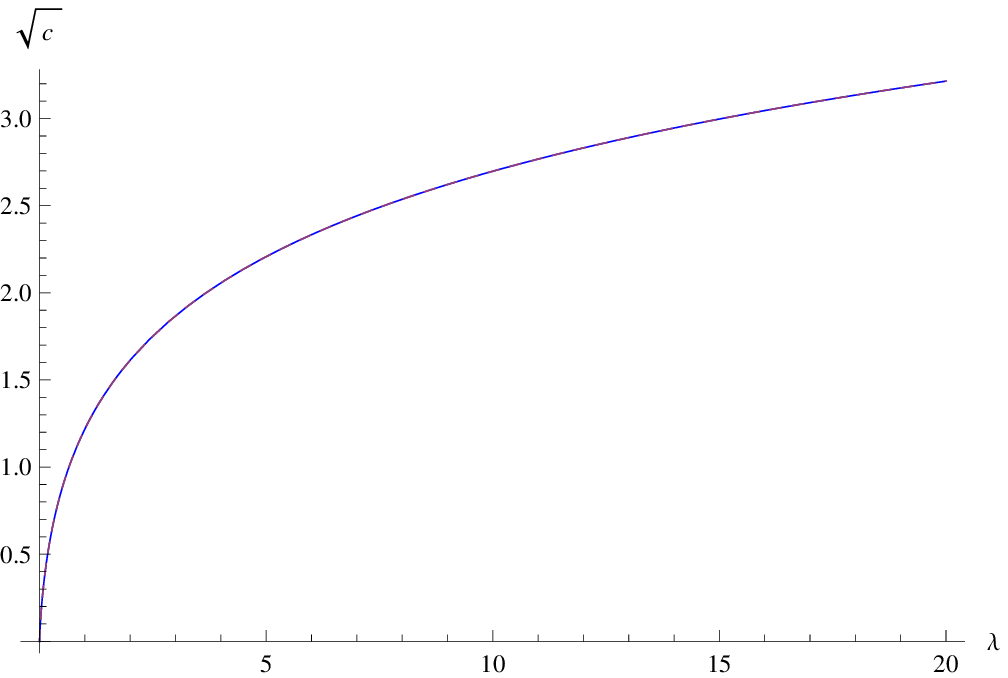}
\label{sscw1}
  }
  \qquad\qquad
  \subfigure[]{\includegraphics[scale=.5]{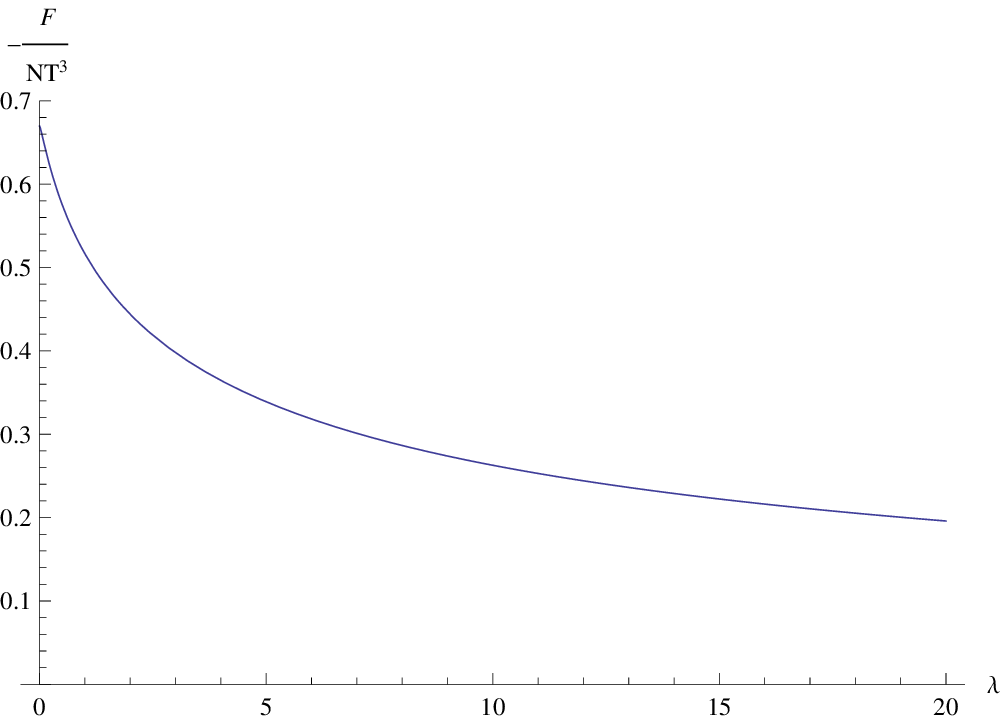}
\label{freenw1}
  }
  \end{center}
  \vspace{-0.5cm}
  \caption{ In Fig.\ref{sscw1}, $\sqrt c$ and $\sqrt\sigma$ are shown as a function of $\lambda$ by undotted and dotted lines, respectively, when $\cN=2$. They are degenerate and finite in all region of $\lambda$. In Fig.\ref{freenw1}, the free energy density normalized by $-NT^3$ is also finite in all region of $\lambda$.}
  \label{w1}
\end{figure}
As seen  from Figure \ref{sscw1}, \ref{freenw1}, the parameters  $\sqrt c, \sqrt \sigma$ and the free energy density are all finite in the whole region of $|\lambda| \in [0,\infty)$. 
This is to be  contrasted with the case of a fermion coupled to the CS
 gauge field \cite{Giombi:2011kc} where the thermal mass diverges and the free energy density vanishes at $|\lambda|=1$
We will comment on this difference more in Section \ref{comments}.

The asymptotic behaviors as $ |\lambda| \rightarrow 0, \infty$ are also worth commenting on. 
Around $|\lambda| \sim 0$ we see from \eqref{scsusy} and \eqref{fesusy1} that $\sqrt{c}$ and the free energy behaves as 
\beal{
\sqrt c &\sim  -2 \lambda \log \lambda (- \log\lambda)^{1 \over \log\lambda},
\label{sqrtcsmall}\\
F&\sim -{NT^3 \over 6\pi}  \biggl( {21 \over 2} \zeta(3) +8 \lambda^2(- \log\lambda)^{3 \over \log\lambda} (\log\lambda)^2 (3 \log2 (1+\log\lambda) -\log \lambda)  \biggr).
}
Around $|\lambda| \sim \infty$, $\sqrt{c}$ diverges and the free energy density vanishes as
\beal{
\sqrt c &\sim \log 4\lambda - \log \log 4 \lambda + {\log \log 4 \lambda \over \log 4\lambda},\\
F&\sim -\frac{N T^3}{6\pi} \frac{(\log(4\lambda))^3}{\lambda}.
}

\subsubsection{${\cal N}=1$ case}
Another interesting case to consider is that of ${\cal N}=1$ SUSY\footnote{ 
Application to $\cN=1$ is first suggested by S. Kim. We thank him for discussion.
Issues related to conformal invariance 
in this case will be briefly discussed in Section \ref{discussion}}. 
$\cN=1$ supersymmetry is realized when the coefficients \eqref{coefficients} are given by
\be
\label{defwsu}
x_4={1+w \over 2}, \, x_4'=w, \, x_4''=w-1, \, x_6=w^2, 
\ee 
where $w$ is a constant, which appears as a coefficient of superpotential \eqref{superpot}. 
 More details on the   Chern-Simons-matter action  of this theory can be found in  Appendix \ref{susycsn1}. 
The large $N$ limit of this theory is therefore characterized by two parameters, $\lambda$ and $w$. 

Let us now briefly consider the behavior for varying values of $w \in [0,1]$. 
\paragraph{${\bf w=0}$}

\begin{figure}
   \begin{center}
   \subfigure[]{\includegraphics[scale=.5]{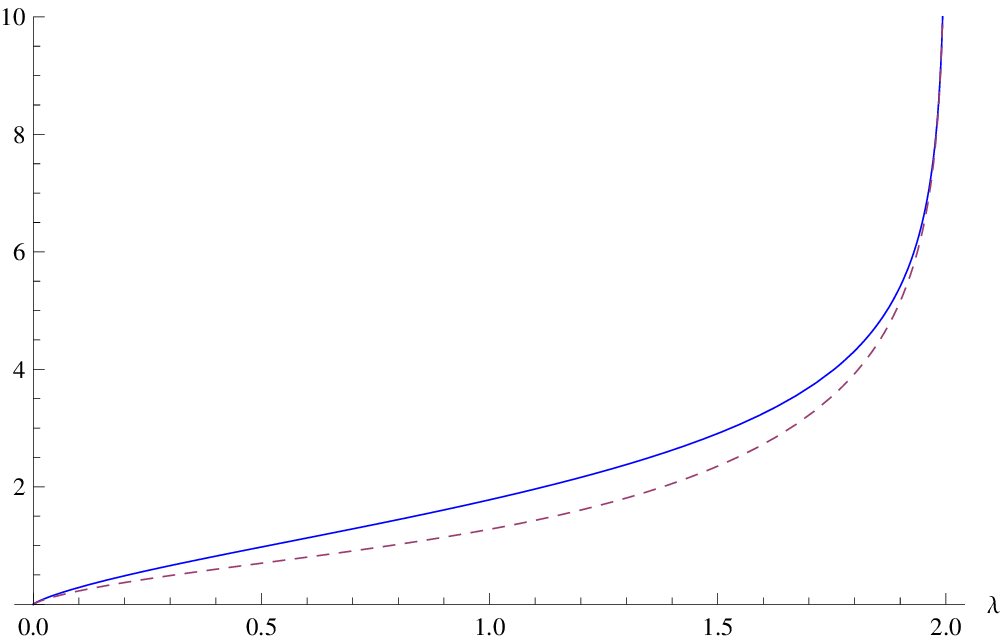}
\label{sscw0}
  }
  \qquad\qquad
  \subfigure[]{\includegraphics[scale=.5]{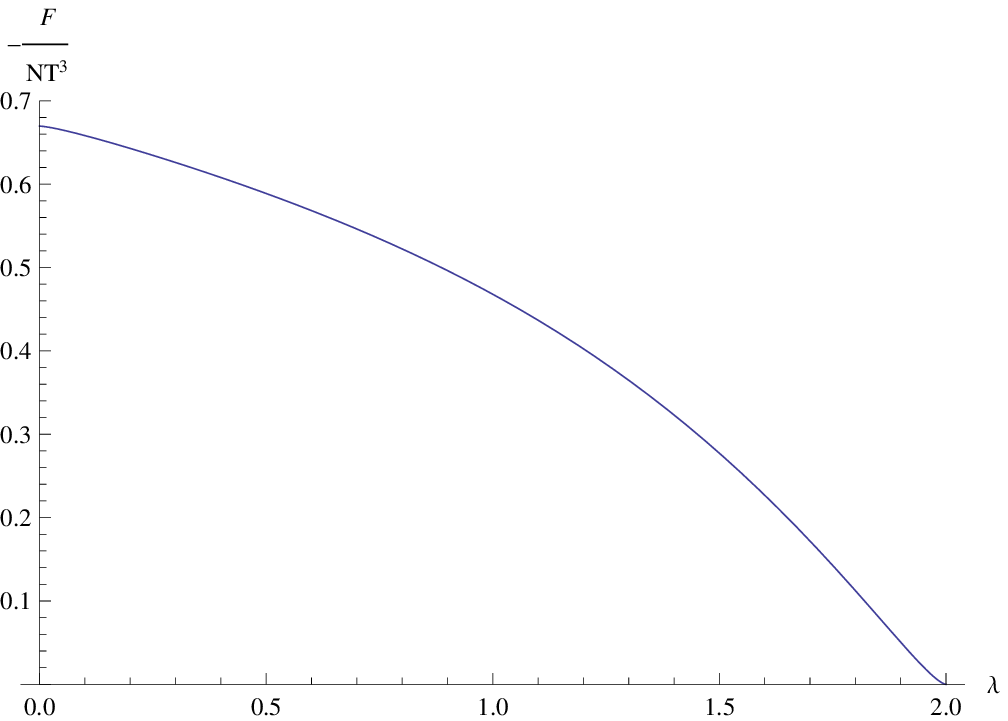}
\label{freenw0}
  }
  \end{center}
  \vspace{-0.5cm}
  \caption{ In Fig.\ref{sscw0}, we plot the $\sqrt c$ and $\sqrt\sigma$ as a function of $\lambda$ by undotted and dotted lines, respectively, when $w=0$. They are finite when $0\leq \lambda < 2$ and divergent at $\lambda=2$. In Fig.\ref{freenw0}, the free energy density divided by $-NT^3$ is shown as a function of $\lambda$ with $w=0$. $-F>0$ when $0\leq \lambda < 2$ and $F=0$ at $\lambda=2$.}
  \label{w0}
\end{figure}

In this  case the  $\cN=1$ superpotential \eqref{superpot} vanishes. 
The graphs of $\sqrt c, \sqrt\sigma$ are shown in Fig.\ref{sscw0} and that of free energy density (with some normalization) is in Fig.\ref{freenw0}. 

From Fig.\ref{sscw0}, one can read off that $\sqrt c$ and $\sqrt\sigma$ monotonically increase up to $\lambda=2$ and are 
divergent at that point. 
Since $\sqrt c$ and $\sqrt\sigma$  correspond to the thermal masses of the fermion and  scalar respectively 
we see that these masses diverge at $\lambda=2$. 
Correspondingly the free energy density with minus sign, which is shown in Fig.\ref{freenw0}, is a monotonically 
decreasing function of $\lambda$ and vanishes at $\lambda=2$.  This behavior of the system is consistent with the following picture. As the coupling constant $\lambda$ increases in value the thermal masses grow and the finite temperature
correlations become   shorter ranged. Finally,  at  $\lambda=2$ the thermal masses diverge
causing the effective degrees of freedom to vanish and thus leading to a vanishing free energy density.

\begin{figure}
  \begin{center}
  \subfigure[]{\includegraphics[scale=.5]{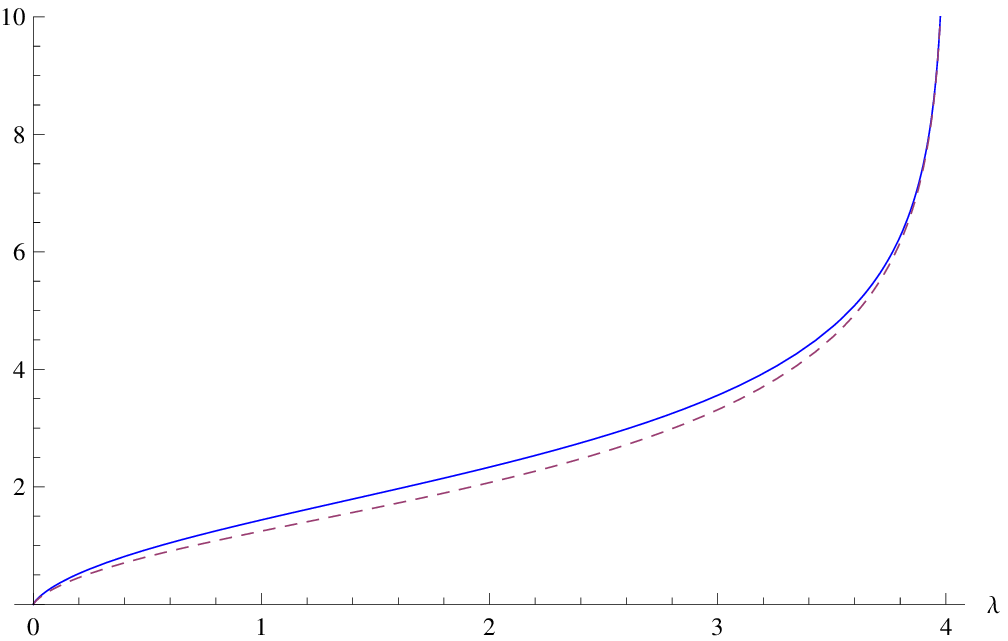}
\label{sscw0.5}
  }
  \qquad\qquad
  \subfigure[]{\includegraphics[scale=.5]{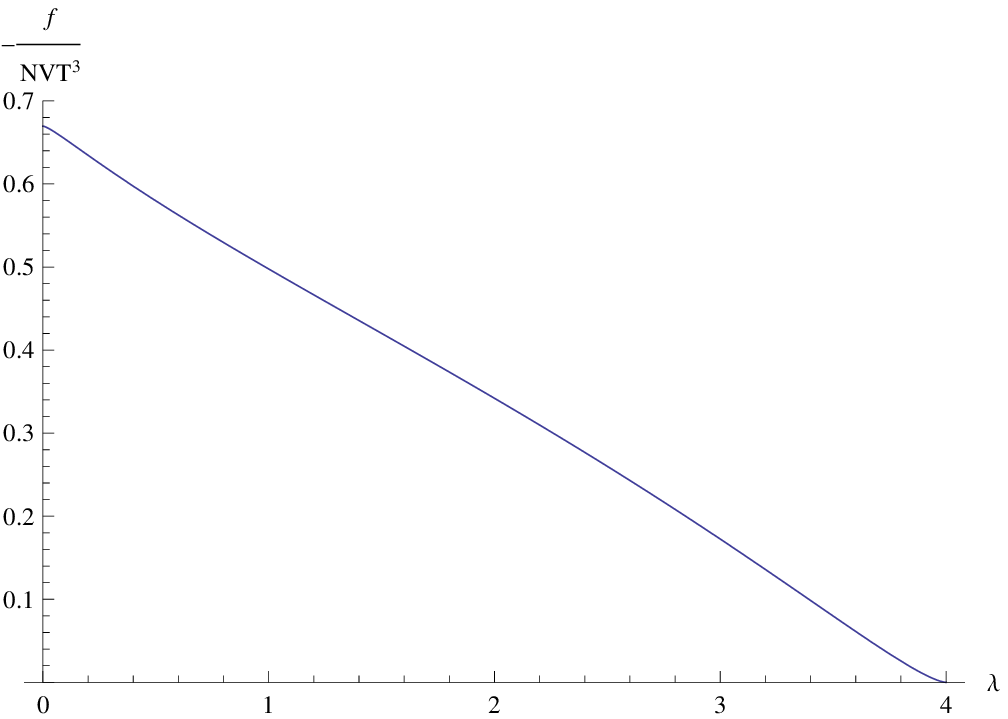}
\label{freenw0.5}
  }
  \end{center}
  \vspace{-0.5cm}
  \caption{ In Fig.\ref{sscw0.5}, $\sqrt c$ and $\sqrt\sigma$ are plotted as a function of $\lambda$ by undotted and dotted lines, respectively, when $w=0.5$. They are finite when $0\leq \lambda < 4$ and divergent at $\lambda=4$. In Fig.\ref{freenw0.5}, the free energy density divided by $-NT^3$ is shown. 
$-F>0$ when $0\leq \lambda < 4$ and $F=0$ at $\lambda=4$.}
  \label{w0.5}
\end{figure}

\paragraph{${\bf 0 \leq w \le 1}$}
In this region, $\sqrt c$, $\sqrt\sigma$ and the free energy behave  qualitatively in the same manner as the  $w=0$ case. 
Both $\sqrt c$ and $\sqrt\sigma$ monotonically increase with respect to $\lambda$ and blow up at some $\lambda =\lambda_{cr}$, and
 the free energy, with a minus sign,  decreases monotonically in term of $\lambda$ and vanishes 
at $\lambda=\lambda_{cr}$.  
We show representative  graphs of these functions  in Fig.\ref{sscw0.5}, \ref{freenw0.5} when $w=\half$ and 
$\lambda_{cr} =4$. 

The $w=1$ case corresponds to the ${\cal N}=2$ theory. As $w$ approaches  this value
one find that $\lambda_{cr}$ goes to larger values eventually becoming infinite when $w=1$,  in agreement with our discussion of the 
${\cal N}=2$ theory above.

\subsection{Additional comments}
\label{comments}

In this subsection we comment further on the results obtained above for the solution to the gap equations and the free energy at finite temperature.
We saw in the previous subsection that for the ${\cal N}=2$ case the magnitude of the free energy $|F|$ 
  monotonically decreases from its free value and vanishes in the limit when $\lambda\rightarrow \infty$. The thermal masses for the fermion and the boson are also well defined for all finite values of $|\lambda|$ and diverge  at $|\lambda|\rightarrow \infty$. 

This behavior is different from what was found in the study of a single fundamental fermion coupled to the CS gauge field
\cite{Giombi:2011kc}. In that case the free energy vanishes at $|\lambda|=1$ with  the thermal mass also diverges at that 
value of $|\lambda|$. This behavior suggests that the conformal theory stops existing beyond $|\lambda|=1$.
An explanation was given in \cite{Giombi:2011kc} as follows. First, consider  
 a theory in the UV which contains a Yang Mills term in its Lagrangian besides also containing a CS term. The YM term dominates in the UV and this theory can be regulated using any standard regulator. Now starting with such a theory suppose we match in the IR to a theory which only contains the CS term for the gauge field. 
The YM coupling, $g_{YM}^2$ has dimensions of mass so it would natural to do this sufficiently below the $g^2_{UV}$  mass scale.
On carrying out this matching one finds that the level of the CS term in the IR theory is shifted compared to its value in the 
UV theory. This shift was calculated in \cite{Pisarski:1985yj}  and found to be 
\be
|k_{IR}| = |k_{UV}| +N.
\label{cslevelshift}
\ee
In the UV theory, since the UV behavior is that of a conventional YM theory, $k_{UV}$ can take all values
and therefore $|\lambda_{UV}|\in [0,\infty)$.
From \eqref{cslevelshift} it then follows that  $\lambda= {N\over k_{IR}}$ satisfies the relation
\be
|\lambda| = {|\lambda_{UV}| \over 1 + |\lambda_{UV}| } \leq 1, 
\ee
in agreement with the expectation coming from the free energy calculation that the conformal theory does not exist for 
larger values of $|\lambda|$. 

In fact an argument along these lines, now for the ${\cal N}=2$ theory leads to the conclusion that the conformal theory in this case should continue to exist for all values of $\lambda$. In the ${\cal N}=2$ case if one starts in the UV with an additional Yang Mills term for the gauge field one must also add its supersymmetric completion which involves additional  adjoint fermions.
It turns out that when one matches to the IR theory without the SUSY Yang Mills term now there is no shift in the CS level.
The gauge field results in the shift (\ref{cslevelshift}) but the two (real) fermions give rise to an exactly opposite shift
 making the net total shift in the CS level  vanish\footnote{In addition the supersymmetric multiplet also contains a real scalar.
However integrating it out does not lead to a shift in the CS level.}.  This conclusion is in agreement with our results for the ${\cal N}=2$ theory.
As mentioned above in this case the free energy does not vanish for any finite value of $\lambda$ (Figure \ref{freenw1}) and the thermal mass of the 
fermion and boson also remain finite for all finite values of $\lambda$ (Figure \ref{sscw1}). 

Another interesting case is one with ${\cal N}=1$ supersymmetry with the parameter $w$ defined in  (\ref{defwsu}) set to vanish. 
In this case the SUSY completion of the Yang Mills term requires us to add one real adjoint fermion. Integrating out this fermion
results in a shift $k\rightarrow k-N/2$ so that the net shift, after including the effects of the gauge field,  are 
\be
\label{nsn1}
|k_{IR}|=|k_{UV}|+{N \over 2}
\ee
and the resulting value of the 't Hooft coupling is 
\be
|\lambda| = {|\lambda_{UV}| \over 1 + {|\lambda_{UV}| \over 2} } \leq 2.
\ee
When $\lambda_{UV}\rightarrow \infty$ we see that $\lambda\rightarrow 2$ which is exactly the value we found in Figure \ref{sscw0}, \ref{freenw0} where the 
free energy vanishes and the thermal masses blow up. 

These observations  serve as consistency checks on the results we have obtained.%
\footnote{
In the purely Bosonic theory $k_{IR}$ and $k_{UV}$ satisfy the same relation as in the purely Fermionic case,  \eqref{cslevelshift}. This would have suggested that the thermal mass diverges and the Free energy vanishes at $\lambda=1$. However our results in \eqref{sigmasol} and \eqref{fb} when applied to the purely Bosonic theory  show that this does not happen. For this theory the thermal mass depends on the couplings in the combination $\lambda'\equiv \lambda \sqrt{1+3 x_6}$, \eqref{fb}, and one finds that the thermal mass is finite and the Free energy is non vanishing for all values of $\lambda'$.
Thus the behaviour of the  Bosonic theory does not fit within the  general discussion given above. We leave a full understanding of this issue for the future. 
} 
 

\section{Discussion}
\label{discussion}
In this paper we have studied $SU(N)$ level $k$ 
 Chern-Simons gauge theory with matter consisting of fermions and bosons in the fundamental
 representation. We calculated the self-energy for the matter fields and also obtained the free energy at finite temperature,
in the 't Hooft limit. Our leading large $N$ results, which were obtained using a saddle point method, are valid to all orders in the 't Hooft coupling. 
Although our emphasis in the study  has been on the ${\cal N}=2$ theory with one chiral superfield in the 
fundamental representation,  much of the analysis is  general and would apply to other conformal field theories in this class as well. 
Our results show that the free energy, $F$,  of the ${\cal N}=2$ theory is well behaved for all values of the 't Hooft coupling.
Starting from the free theory, at $\lambda=0$, the magnitude of the  free energy, $|F|$, monotonically decreases and 
 finally vanishes at strong coupling,  $|\lambda| \rightarrow \infty$. In addition the thermal mass of the boson and fermion, in the ${\cal N}=2$ theory, also remain finite for all finite $\lambda$. 
The existence of the solution for all values of $\lambda$ is consistent with the exact superconformal symmetry of this system for all values of $N$ and $k$. 
We have discussed the relation of our system to UV theory in Section \ref{comments}.

Our analysis  for the self energy and the free energy can also be applied to theories with ${\cal N}=1$ supersymmetry or 
no supersymmetry, as long as they are conformally invariant. We have not carried out a detailed analysis of when conformal invariance can be preserved in this larger class of theories.  The results in \cite{Avdeev:1992jt} suggest that conformally invariant theories 
with ${\cal N}=1$ or no supersymmetry can indeed exist. We leave a more detailed analysis of this question for the future. 

\paragraph{Lorentz invariance}
Our calculations were based on the techniques developed in \cite{Giombi:2011kc} and were carried out  by choosing 
  light cone gauge $A_-=0$ in the non-thermal directions and by  using dimensional 
regularisation. The choice of gauge we have made is unconventional in Euclidean space and  more study is needed to ensure that
the resulting theory is consistent and Lorentz invariant. 
Some evidence was provided for this in \cite{Giombi:2011kc} especially in the case of the anomalous dimension of the operator $\bar\psi\psi$ which was calculated up to 2-loops in perturbation theory in various gauge. Our calculations which explore a larger class of theories add to 
this evidence. It is worth noting that, in the theories we study, 
the poles in the zero temperature propagators occur at $p^2=0$, which is  Lorentz invariant. 


\paragraph{Higher spin currents}
It is also interesting to study the structure of the higher spin currents in our theory. The higher spin currents for  such theories  with vector matter  have been discussed in \cite{Maldacena:2011jn,Maldacena:2012sf},
and it has been shown that the resulting current algebra is a  very powerful tool  to determining correlation functions in the theory. 
These currents are also important in making a closer connection with dual Vasiliev theories. 

 It is easy to see that in the theory we are studying, 
 with one fermion and one boson in the fundamental representation, and with the double trace Yukawa couplings and the triple 
trace $\phi^6$ coupling, \eqref{generalaction}, there are two  currents for every integer 
spin $s\ge 1$ which are approximately conserved with a
conservation law%
\footnote{There is an exactly conserved current for spin $s=1,2$.}
 schematically  of the form \eqref{conse}.
These currents can be built as follows. We start with say the current of spin $s$ in the free fermion case, then 
replace all ordinary derivatives with covariant derivatives. The effect of the covariant derivatives, which do not commute, 
and the Yukawa and $\phi^6$ terms in the Lagrangian results in the current not being traceless in general.
After adding appropriate multi-trace terms traceless can be restored. This traceless spin $s$ current is not conserved, again 
because of the non-commutation of the covariant derivatives and the presence of the Yukawa and $\phi^6$ couplings. 
The resulting terms  which violate current conservation are of double trace and triple trace type and are 
suppressed by appropriate powers of $N$ as shown in (\ref{conse}).
A similar construction can be carried out starting with the spin $s$ current in the free boson case resulting in the 
two approximately conserved currents for each integer $s \ge 1$ that we mentioned above. 
In addition, this is important in the analysis of  SUSY theories, there are partner currents which have half-integer spin. 

It would be very interesting to explore current equations (analogues of \eqref{conse}) for the ABJ type theories 
\cite{Aharony:2008gk,Chang:2012kt}.
For example in the case of $M=N$ and the adjoint representation of $U(N)$ the gauge invariant currents of higher spin $s\ge 3$ are 
 classically not
conserved. The divergence of the current is not suppressed by powers of $\cO(1/N),$ as can be easily deduced by large $N$ power counting.
This has
the implication that all higher spin ($s\ge 3 $) gauge symmetries of the bulk 
theory  are explicitly broken presumably by boundary conditions.
 

\paragraph{ABJ duality}
Adding more flavors and the exploring theories with more supersymmetry
is interesting. Particularly interesting is the case with ${\cal N}=6$ SUSY \cite{Aharony:2008gk}. 
It has been argued that the $\cN=6$ theory has the duality \cite{Giveon:2008zn} between the theories with gauge groups $U(M+l)_k\times U(M)_{-k}$ and $U(M)_k\times U(M+k-l)_{-k}$.
For the case where $M=1, l=N$ this reduces to a duality between a $U(N+1)_k\times U(1)_{-k}$ theory and a $U(1)_k \times U(1+k-N)_{-k}$ theory. 
The 't Hooft coupling of the non-Abelian groups on the two sides of this duality in the large $N$ limit are $\lambda = {N \over k}$ and $\tilde{\lambda}= 1-\lambda$ respectively. 
The free energy of the two theories at finite temperature should be equal from this duality, once we also account for the change $N \rightarrow k-N$. 
The $U(1)$ theory is in the large flavor limit here.  

We speculate the following two reasons to explain why the free energy formula presented in this paper does not have the duality: 

\begin{enumerate}[i{)}]
\item The presence of an additional $U(1)$ and many flavors in principle leads to modified gap equations. Even though the saddle point of the $SU(N)$ theory continues to solve the new gap equations there are in principle other solutions, whose inclusion in the path integral may restore the duality.

\item The large $N$ saddle point analysis presented in this paper may be rendered invalid due to infrared divergences in sub-leading orders in $1/N$.%
\footnote{We would like to thank Shiraz Minwalla and Ofer Aharony for drawing our attention to this possibility.} 
 However this possibility seems remote because both fermions and bosons have $\cO(1)$ thermal masses.

\item The use of manifestly lorentz invariant gauge may also throw light on this issue. 

\end{enumerate}

Some other interesting problems to pursue include comparing our calculation with the high temperature limit of the theory on $\bS^2\times \bS^1$ where a Gross-Witten-Wadia type phase transition \cite{Gross:1980he,Wadia:1980cp} is expected at a (dimensionless temperature) $T \sim N^{1/2}$ \cite{Giombi:2011kc}. 
It will also be interesting to calculate the partition function of the $SU(N)$ theory on $\bS^3$. 
Other gauge groups are also interesting to consider, especially in the case with $Sp(N)$ gauge symmetry in the context of dS/CFT in \cite{Strominger:2001pn,Anninos:2011ui,Ng:2012xp}. This has been independently suggested by \cite{Das:2012dt}. Finally, it would be worth exploring connections with Vasiliev theories in more detail.

\section*{Acknowledgments} 
We thank Ofer Aharony, Rajesh Gopakumar, Seok Kim, Sunil Mukhi and Shiroman Prakash for discussions. 
We are grateful to  Chi-Ming Chang, Nilanjan Sirkar and Xi Yin for initial collaboration and discussions.
We especially thank Shiraz Minwalla for extensive collaboration on many aspects of this work and for critical discussions. 
We are grateful to Diptimoy Ghosh for help with the figures. 
SRW and SY thank the Isaac Newton Institute for Mathematical Sciences for hospitality, where this work was partly done during the program `Mathematics and Applications of Branes in String and M-theory'.
SPT's research is supported by a J. C. Bose fellowship. 
We would like to thank the people of India for their generous support to research in basic sciences.

\appendix
\section{Conventions}
\label{convention}

In this appendix, we collect our convention used in this paper. 

\paragraph{Normalization on gauge group}
The $SU(N)$ gauge field $A_\mu$ will be expanded by the generators of the gauge group $T^a$ as $A_\mu=\sum_{a} A_\mu^a T_a$, where $a$ runs from $1$ to $N^2-1$. $T_a$ is normalized so that $\tr (T^a T^b)=\delta^{ab}$.
Note that
\be
\sum_{a}  (T^a)_m^n (T^a)_p^q= 
\delta_m^q \delta_p^n - {1 \over N} \delta_m^n \delta^p_q.
\label{tata}
\ee

\paragraph{Spinor contraction}
We contract spinor indices for fermion bilinear in the following way. 
\be
\psi A \chi := \psi_\alpha A^\alpha{}_\beta \chi^\beta = \varepsilon_{\alpha\gamma}  \psi^\gamma A^\alpha{}_\beta \chi^\beta,
\ee
where $A$ is arbitrary two by two matrix and $\varepsilon_{\alpha\beta}$ is $\varepsilon$-invariant tensor.

\paragraph{Gamma matrices (Euclidean)}
We choose the gamma matrices $\gamma^\mu$ as Pauli matrices.
\begin{equation}
 \g^1=\sigma^1 = \left(
\begin{array}{cc}
 0 &1 \\
 1&0
\end{array}\right) ~~;~~
\g^2=\sigma^2 = \left(
\begin{array}{cc}
 0 & -i\\
 i & 0
\end{array}\right) ~~;~~
\g^3=\sigma^3 =\left(
\begin{array}{cc}
 1 &0 \\
 0&-1
\end{array}\right).
\end{equation}

\paragraph{Coordinates}
In the main text, we have considered Lorentzian space-time $\bR^{1,1}$ from \eqref{gauge} to \eqref{integratedaction}. 
$x^{\pm}, A_{\mp}, \partial_{\mp}, p_{\mp}$ represent
\begin{eqnarray}
x^{\pm} = \f{x^1 \pm  x^2}{\sqrt{2}},\quad
A_{\mp} = \f{A_1 \pm  A_2}{\sqrt{2}},\quad 
\partial_{\mp} = \f{\partial_1 \pm  \partial_2}{\sqrt{2}},\quad 
p_{\mp} = \f{p_1 \pm p_2}{\sqrt{2}}.
\eea
After \eqref{integratedaction}, we consider Euclidean space $\bR^2$ and $x^{\pm}, p_{\mp}$ represent 
\begin{eqnarray}
x^{\pm} = \f{x^1 \pm i x^2}{\sqrt{2}},\quad
p_{\mp} = \f{p_1 \pm i p_2}{\sqrt{2}}.
\eea
We denote the absolute value of 2-plane momentum as $p_s$. 
That is, 
\bea
p_s^2 := p_1^2+p_2^2 = 2 p_+ p_-.
\end{eqnarray}

\paragraph{Fourier expansion}
For a complex field $\Phi(x)$, we can define $\Phi(p)$ by Fourier expansion. On $\bR^3$, $\Phi(x)$ can be expanded as
\be
\Phi (x) = \int \f{d^3 p}{(2 \pi)^3} e^{i p x} \Phi(p). 
\ee
On $\bR^2\times \bS^1$, $\Phi(x)$ can be expanded as
\be
\Phi (x) = {1 \over \beta} \sum_{p_3} \int \f{d^2 p}{(2 \pi)^2} e^{i p x} \Phi(p), 
\ee
where $p_3=p_{3,n}$ is determined by boundary condition of the field $\Phi$ with respect to $\bS^1$.

\paragraph{Gaussian integration}
The Gaussian integration in momentum space is done as follows. 
For a complex scalar field and a positive function with respect to momentum $a(p)$, 
\be
\int \cD \phi \cD\bar\phi e^{-\int \f{d^3 p}{(2 \pi)^3} \bar\phi(-p) a(p) \phi(p) }
= (\mathrm{Det} a(p) )^{-1}
= e^{- V \int \f{d^3 p}{(2 \pi)^3} \log a(p) }, 
\ee 
where we set $V= (2 \pi)^3 \delta^3(0)$. 
For a fermionic field and a positive-definite two by two matrix $M(p)$, 
\be
\int \cD \psi \cD\bar\psi e^{-\int \f{d^3 p}{(2 \pi)^3} \bar\psi(-p) M(p) \psi(p) }
= \mathrm{Det} M(p) 
= e^{V \int \f{d^3 p}{(2 \pi)^3} \log\det M(p) }.
\ee 

\section{Schwinger-Dyson equation}\label{SD}
In what follows we shall derive Schwinger-Dyson equations for scalar and fermion field and confirm that they reproduce the gap equations obtained as a saddle point of the exact effective action. We will follow \cite{Wadia:1980rb}.

\paragraph{Schwinger-Dyson equation for scalar fields}
First we derive Schwinger-Dyson equation for the scalar field. 
The Schwinger-Dyson equation can be derived via
\begin{eqnarray}
 0 = \int D\phi D \bar{\phi} \f{\delta}{\delta \bar{\phi}^m (-p)} \left( e^{-S_1} \bar{\phi}^n (q)\right), 
\end{eqnarray}
where $S_1$ is given by \eqref{integratedaction}.
Since we are interested in gauge singlet sector, 
we contract the gauge index. 
\begin{eqnarray}
 0 ={1 \over Z} \int D\phi D \bar{\phi} \sum_m 
\f{\delta}{\delta \bar{\phi}^m (-p)} \left( e^{-S_1} \bar{\phi}^m (q)\right).
\end{eqnarray}
Here we divide both sides by the total partition function.
From direct calculation, 
the right-hand side becomes
\beal{ 
\mbox{r.h.s.} &= N \delta^3(q+p)
- \f{p^2}{(2 \pi)^3} N \langle \chi(q+p, {-q+p \over 2}) \rangle\nonumber\\
& - \frac{2^3 }{(2\pi)^3} N\int \f{d^3 q_1}{(2 \pi)^3} \f{d^3 q_2}{(2 \pi)^3} 
\left(C_1(2q_1 -2p, q_1, q_2) + C_1(-2q_1+2p, q_2, q_1) \right) \times \nn
& \qquad
 \aver{\chi(q+2q_1-p,{ -q+2q_1 - p \over 2})  \chi(-2q_1+p,q_2)}\nonumber \\
&  - \frac{2^3 }{(2\pi)^3} N
\int \f{d^3 P}{(2 \pi)^3}\f{d^3 q_1}{(2 \pi)^3}\f{d^3 q_2}{(2 \pi)^3}\f{d^3 q_3}{(2 \pi)^3} \nn 
& \qquad \Big\lbrack C_2(2q_1-2p,P,q_1,q_2,q_3)+C_2(P,2q_1-2p,q_2,q_1,q_3)\nonumber +C_2(2p-2q_1-P,P,q_3,q_2,q_1)\Big\rbrack \times \nn 
& \qquad\aver{ \chi(q + 2q_1-p ,{-q+2p-p \over 2}) \chi(P,q_2)  \chi(-2q_1+2p-P,q_3)}\nonumber\\
&-2\lambda_{4}N^2 \frac{2^3 }{(2\pi)^3} 
\int \f{d^3 q_1}{(2 \pi)^3} \f{d^3 q_2}{(2 \pi)^3} 
\aver{ \xi_{I}(-2q_2+p,q_1)\chi(q+2q_2-p, {-q+2q_2-p \over 2}) }.
}
In the large $N$ limit the leading term is dominated by factorized correlators.
\beal{ 
\mbox{r.h.s.} &= N \biggl[ \delta^3(q+p)
- \f{p^2}{(2 \pi)^3} \langle \chi(q+p, {-q+p \over 2}) \rangle\nonumber\\
& - \frac{2^3 }{(2\pi)^3} \int \f{d^3 q_1}{(2 \pi)^3} \f{d^3 q_2}{(2 \pi)^3} 
\left(C_1(2q_1 -2p, q_1, q_2) + C_1(-2q_1+2p, q_2, q_1) \right) \times \nn
& \qquad
 \aver{\chi(q+2q_1-p, { -q+2q_1 - p \over 2})}\aver{  \chi(-2q_1+p,q_2)}\nonumber \\
&  - \frac{2^3 }{(2\pi)^3} \int \f{d^3 P}{(2 \pi)^3}\f{d^3 q_1}{(2 \pi)^3}\f{d^3 q_2}{(2 \pi)^3}\f{d^3 q_3}{(2 \pi)^3} \nn 
& \qquad \Big\lbrack C_2(2q_1-2p,P,q_1,q_2,q_3)+C_2(P,2q_1-2p,q_2,q_1,q_3)\nonumber +C_2(2p-2q_1-P,P,q_3,q_2,q_1)\Big\rbrack \times \nn 
& \qquad\aver{ \chi(q + 2q_1-p ,{-q+2p-p \over 2})}\aver{ \chi(P,q_2) }\aver{ \chi(-2q_1+2p-P,q_3)}\nonumber\\
&-2\lambda_{4}N^2 \frac{2^3 }{(2\pi)^3} 
\int \f{d^3 q_1}{(2 \pi)^3} \f{d^3 q_2}{(2 \pi)^3} 
\aver{ \xi_{I}(-2q_2+p,q_1)} \aver{\chi(q+2q_2-p, {-q+2q_2-p \over 2}) }+ \cO({1 \over N}) \biggr].
}
From \eqref{unitinsertion}, \eqref{condm}, \eqref{pis}, $\aver{\chi(P,q)}$ and $\aver{\xi(P,q)}$ are given by 
\beal{
 \aver{ \chi(P,q)} &= \f{1}{q^2+\Sigma_{B}(q)} (2\pi)^3 \delta^3(P),\quad
\aver{\xi(P,q)} =-{ 1 \over i\gamma^\mu q_\mu +\Sigma_F(q) }(2\pi)^3\delta^3(P).
}
Using the above ansatz we can proceed
\beal{ 
\mbox{r.h.s.} &= {N \delta^3(q+p) \over q^2 + \Sigma_B(q)} 
\biggl[ \Sigma_B(q) - 
\int \f{d^3 q_2}{(2 \pi)^3}\f{d^3 q_3}{(2 \pi)^3} 
\Big\lbrack C_2(q,q_2,q_3)+C_2(q_2,q,q_3)\nonumber +C_2(q_3,q_2,q)\Big\rbrack \times \nn 
& \qquad {1 \over q_2^2 + \Sigma_B(q_2)} {1 \over q_3^2 + \Sigma_B(q_3)} + \lambda_{4}N \int \f{d^3 q'}{(2 \pi)^3} \tr( {1 \over i\gamma^\mu q'_\mu + \Sigma_F(q')} ) \biggr]. 
}
Note that the quadratic term with respect to $\chi$ vanishes and $C_2(p,q,r)$ is given by \eqref{c2}. 
Since this was required to vanish we obtain Schwinger-Dyson equation as
\beal{
\Sigma_B(q) &=
\int \f{d^3 q_2}{(2 \pi)^3}\f{d^3 q_3}{(2 \pi)^3} 
\Big\lbrack C_2(q,q_2,q_3)+C_2(q_2,q,q_3)\nonumber +C_2(q_3,q_2,q)\Big\rbrack  {1 \over q_2^2 + \Sigma_B(q_2)} {1 \over q_3^2 + \Sigma_B(q_3)} \nn 
& \qquad - \lambda_{4}N \int \f{d^3 q'}{(2 \pi)^3} \tr( {1 \over i\gamma^\mu q'_\mu + \Sigma_F(q')} ).
}
This is the same as the gap equation of the scalar field \eqref{gapeqnall}.

\paragraph{Schwinger-Dyson equation for fermionic field}
We repeat the same procedure for the fermionic field. 
The starting point is the following identity. 
\begin{eqnarray}
 0 = \int D\psi D \bar{\psi} \f{\delta}{\delta \bar{\psi}^m (-p)} \left( e^{-S_1} \bar{\psi}^n (p')\right), 
\end{eqnarray}
or by taking contraction for gauge index 
\begin{eqnarray}
 0 ={1 \over Z} \int D\psi D \bar{\psi} \sum_m 
\f{\delta}{\delta \bar{\psi}^m (-p)} \left( e^{-S_1} \bar{\psi}^m (p')\right).
\end{eqnarray}
In purely fermionic case this calculation was already done in \cite{Giombi:2011kc}. The difference between their case and our case is one term coming from $S_{BF}$. The equation (2.27) in \cite{Giombi:2011kc} is corrected by a constant.
\begin{eqnarray}
N  (2 \pi)^3 \delta^3(p'+p)&=&ip_{\mu}\gamma^{\mu} \langle \psi_m(p) \bar{\psi}^m (p')\rangle\nonumber\\
&+&\frac{2\pi i}{k}\int \frac{d^3 r}{(2\pi)^3}\f{d^{3}q}{(2\pi)^3}\frac{1}{q^{+}}\gamma^{+}\langle\psi_{n}(p-q)\bar{\psi}^{n}(-r)\gamma^{3}\psi_{m}(r+q)\bar{\psi}^{n}(p')\rangle\nonumber\\
&-&\frac{2\pi i}{k}\int \frac{d^3 r}{(2\pi)^3}\f{d^{3}q}{(2\pi)^3}\frac{1}{q^{+}}\gamma^{3}\langle\psi_{n}(p-q)\bar{\psi}^{n}(-r)\gamma^{+}\psi_{m}(r+q)\bar{\psi}^{n}(p')\rangle\nonumber\\
&+& \lambda_{4}N\int \frac{d^3 q}{(2\pi)^3}\frac{1}{ip^{'}_{\mu}\gamma^{\mu}+\Sigma_{F}(p')}\aver{\bar{\phi}(p^{'}+p-q)\phi(q)}.
\end{eqnarray}
In the large $N$ limit the factorization leads to
\begin{eqnarray}
  \langle \psi_m(p) \bar{\psi}^m (p') \rangle &=&\frac{N}{i\gamma^{\mu}p_{\mu}}(2 \pi)^3 \delta^3(p'+p)\nonumber\\
&-&\frac{1}{i\gamma^{\mu}p_{\mu}}\frac{2\pi i}{k}\int \frac{d^3 r}{(2\pi)^3}\f{d^{3}q}{(2\pi)^3}\frac{1}{q^{+}}\gamma^{+}\langle\psi_{a}(p-q)\bar{\psi}^{a}(-r)\rangle\langle\gamma^{3}\psi_{m}(r+q)\bar{\psi}^{n}(p')\rangle\nonumber\\
&+&\frac{1}{i\gamma^{\mu}p_{\mu}}\frac{2\pi i}{k}\int \frac{d^3 r}{(2\pi)^3}\f{d^{3}q}{(2\pi)^3}\frac{1}{q^{+}}\gamma^{3}\langle\psi_{a}(p-q)\bar{\psi}^{a}(-r)\rangle\gamma^{+}\langle\psi_{m}(r+q)\bar{\psi}^{n}(p')\rangle\nonumber\\
&-&\frac{1}{i\gamma^{\mu}p_{\mu}} \lambda_{4}N\int \frac{d^3 q}{(2\pi)^3}\frac{1}{ip^{'}_{\mu}\gamma^{\mu}+\Sigma_{F}(p')}\aver{\bar{\phi}(p^{'}+p-q)\phi(q)}.
\end{eqnarray}
Plugging the fermionic and bosonic propagator into the above gives the fermion self energy to be 
\begin{eqnarray}
\Sigma_{F}(p)& =& -2\pi i\lambda\int\frac{d^{3}q}{(2\pi)^3}\Bigg(\gamma^{3}\frac{1}{i\gamma^{\mu}q_{\mu}+\Sigma_{F}}\gamma^{+}-\gamma^{+}\frac{1}{i\gamma^{\mu}q_{\mu}+\Sigma_{F}}\gamma^{3}\Bigg)\frac{1}{(p-q)_{-}}\nonumber\\
&+&\lambda_{4}N\int\frac{d^{p}}{(2\pi)^3}\frac{1}{p^2+\Sigma_{B}(p)},
\end{eqnarray}
which is the same as the gap equation for the fermionic field \eqref{gapeqnall}.

\section{On regularization} 
\label{regularization}

In this appendix we have a comment on the dimensional regularization we used in this paper. In our analysis, we encounter momentum integrals which includes UV divergence. They are of the form 
\be
\int \f{d^3 p}{(2 \pi)^3} F(p_s, |p|) \quad \mbox{or} \quad
{1\over \beta} \sum_{p_3} \int \f{d^2 p}{(2 \pi)^2} F(p_s, |p|),
\ee
where $F(p_s, |p|)$ is a function of $p_s = \sqrt{p_1^2+p_2^2}$ and $|p|=\sqrt{p_1^2+p_2^2+p_3^2}$. 
To regularize UV divergence, what we do is to change the dimension from $3$ to $D=3-\epsilon$ in the integration and also momentum according to it. 
\be
\int \f{d^D \hat p}{(2 \pi)^D} F(p_s, |\hat p|) \quad \mbox{or} \quad
{1\over \beta} \sum_{p_3} \int \f{d^{2-\epsilon} p}{(2 \pi)^{2-\epsilon}} F(p_s, |\hat p|),
\ee
where $|\hat p|=\sqrt{p_1^2+p_2^2+\cdots+p_D^2}$. 
It is remarkable that we do not touch the part of $2$-momentum $p_s$ under this process. It is clear that the UV divergence is cured if $\epsilon$ is greater that some positive number. Performing integration, we obtain some function of $\epsilon$. 
Analytically continuing from some positive number $\epsilon$ to 0, 
we obtain what we wanted to evaluate.

In case where the integral includes IR divergence, it can be cured by adding a virtual mass parameter in 2-plane by changing $p_s \to \sqrt{p_s^2+m^2}$ and setting $m=0$ after evaluating the integral.

\section{Detailed calculation}

In this appendix we collect results which will be obtained after detailed calculation. In this appendix and the main text we will use the following integration formula without mentioning
\begin{eqnarray}
\int \f{d^D\hat q}{(2\pi)^D}\f{1}{(\hat q^2+M^2)^a} &=& \f{1}{(4 \pi)^{\f{D}{2}}} \f{\Gamma(a-\f{D}{2})}{\Gamma(a)} \f{1}{M^{2a-D}}, 
\label{formula}
\end{eqnarray}
which is available in a standard textbook of quantum field theory. 

\subsection{$\Delta S_{BF}$}
\label{sbf}
$\Delta S_{BF}(\eta,\bar{\eta})$ is given by 
\beal{
\Delta S_{BF}(\eta,\bar{\eta})&=N \int  \f{d^3P}{(2 \pi)^3} \f{d^3q_1}{(2 \pi)^3} \f{d^3q_2}{(2 \pi)^3} \biggl( 
{2 \pi N \over k} \f{(P+q_1+q_2)_{3}}{(q_1-q_2)_{-}} \{\bar{\eta}(P,q_1) \g_{-} \eta(-P,q_2)-\bar{\eta}(P,q_1) \g_{3} \eta(-P,q_2) \} \nn
&+N \lambda_4' \bar\eta(P,q_1) \eta(-P,q_2)
+N \lambda_4'' \{ \eta(P,q_1) \eta(-P,q_2) + \bar\eta(P,q_1) \bar\eta(-P,q_2)\} \biggr) \nn
&+ N \int \f{d^3P_1}{(2 \pi)^3} \f{d^3P_2}{(2 \pi)^3}  \f{d^3q_1}{(2 \pi)^3} \f{d^3q_2}{(2 \pi)^3}\f{d^3q_3}{(2 \pi)^3}~ { 2 \pi^2 N^2 \over k^2} \times\nn
&\biggl(\tr\left(\xi(P_1,q_1)\gamma_{-} \eta(P_2,q_2)\bar{\eta}(-P_1-P_2,q_3)\g_{-}- \bar{\eta}(P_1,q_1) \gamma_{-} \xi(P_2,q_2)\g_{-}\eta(-P_1-P_2,q_3)\right)\nn
&+ i (P_2-P_1-2q_1-2q_2-4q_3)_{-} \bar{\eta}(P_1,q_1)\g_{-} \eta(P_2,q_2) \chi(-P_1-P_2,q_3)\biggr).
}

\subsection{Check of the solution of the gap equation}
\label{cgesol1}

In this appendix we confirm that the solution presented in Section \ref{solution} satisfies the gap equation \eqref{gapeqnall}. That is, the equation we should like to check is
\begin{eqnarray}~\label{DSeq0}
0 &=& 3N^2 \lambda_6 \int \f{d^3q}{(2\pi)^3}\f{d^3q'}{(2\pi)^3}\f{1}{q^2}\f{1}{q'^2} \nonumber \\
&+&  8 \pi^2 \la^2 \int \f{d^3q}{(2\pi)^3}\f{d^3q'}{(2\pi)^3}
\Big\lbrack \f{(p+q')_{-}(q+q')_{-}}{(p-q')_{-} (q-q')_{-}}\Big\rbrack \f{1}{q^2}\f{1}{q'^2}\nonumber \\
&+& 4 \pi^2 \la^2\int \f{d^3q}{(2\pi)^3}\f{d^3q'}{(2\pi)^3}
\Big\lbrack\f{(p+q)_{-}(p+q')_{-}}{(p-q)_{-} (p-q')_{-}}\Big\rbrack\f{1}{q^2}\f{1}{q'^2}\nonumber\\
&-&N \lambda_{4}\int\frac{d^3 q}{(2\pi)^3}\tr\frac{1}{iq^{\mu}\gamma_{\mu}+\Sigma_{F}(q)}.
\end{eqnarray}
We compute the integrals by dimensional regularization explained in Appendix \ref{regularization}. 

First we focus on the first term, which is given by $3N^2 \lambda_6 J_1^2$ with $M=0$, where 
\begin{eqnarray}
J_1=\int \f{d^D\hat q}{(2\pi)^D}\f{1}{\hat q^2+M^2}. 
\label{j1}
\end{eqnarray}
By using \eqref{formula} we can evaluate this as 
\begin{eqnarray}
J_1= \f{\Gamma(1-\f{D}{2})}{(4 \pi)^{\f{D}{2}}} \f{1}{M^{2-D}} = - {M \over 4\pi} +\cO(\epsilon), \label{j2}
\end{eqnarray}
where $D$ is greater than two for the integration to be convergent. 
$J_1 \to  - {M \over 4\pi}$ after analytic continuation with $\epsilon\to0$. 
Since $J_1$ vanishes under $M=0$, 
the first term also does. 

Next we compute the second and third integrals. 
We consider the second integral, $8\pi^2\lambda^2 J_2$, where 
\begin{eqnarray}
 J_2=\int \f{d^2q}{(2\pi)^2}\f{d^2q'}{(2\pi)^2}
\Big\lbrack \f{(p+q')_{-}(q+q')_{-}}{(p-q')_{-} (q-q')_{-}}\Big\rbrack 
\int \f{d^{(1-\epsilon)}\hat q_3}{(2 \pi)^{1-\epsilon}} \f{1}{q_s^2+\hat q_3^2}
\int \f{d^{(1-\epsilon)} \hat q_3'}{(2 \pi)^{1-\epsilon}} \f{1}{q_s^{'2}+\hat q_3^{'2}}.
\end{eqnarray}
Here we separate out the two light cone directions and remaining $1-\epsilon$ directions. Let us perform the integration in $1-\epsilon$ dimensions. 
\begin{eqnarray}
 J_2 &=& A^{2} \int \f{d^2q}{(2\pi)^2}\f{d^2q'}{(2\pi)^2}
\Big\lbrack \f{(p+q')_{-}(q+q')_{-}}{(p-q')_{-} (q-q')_{-}}\Big\rbrack 
\f{1}{|q_s|^{1+\epsilon}} \f{1}{|q_s'|^{1+\epsilon}}, 
\end{eqnarray}
where we set
\be
A:= \f{\Gamma(1-\f{1-\epsilon}{2})}{(4 \pi)^{\f{1-\epsilon}{2}}}
=\f{1}{2} +\mathcal{O}(\epsilon). 
\ee
To carry out the remaining integral, we use the following formula
\be
\int \f{d^2q}{(2\pi)^2} \f{(q+r)_{-}}{(q-r)_{-}} \f{1}{|q_s|^a} =
- {1 \over \pi(2- a) |r_s|^{a-2} },
\ee
where $a$ is suitably greater than a positive number to cut the UV divergence.
This can be obtained in a similar manner to the formulas (2.17-19) derived in \cite{Giombi:2011kc}. 
Employing the formula twice we find 
\begin{eqnarray}
J_2 
&=&A^{2} \int \f{d^2q'}{(2\pi)^2}
 \f{(p+q')_{-}}{(p-q')_{-}} \f{1}{|q_s'|^{1+\epsilon}} \times 
\left(- {1 \over \pi(1-\epsilon) |q'_s|^{\epsilon-1} } \right) \nn
&=&\left(- {A^{2} \over \pi(1-\epsilon) } \right) \int \f{d^2q'}{(2\pi)^2}
 \f{(p+q')_{-}}{(p-q')_{-}} \f{1}{|q_s'|^{2\epsilon}}  \nn
&=&\left(- {A^{2} \over \pi(1-\epsilon) } \right) \times 
\left( {1 \over \pi(2-2\epsilon) |p_s|^{2\epsilon-2} } \right)  \nn
&=& -\f{p_s^2}{8 \pi^2} +\mathcal{O}(\epsilon).
\end{eqnarray}
Similarly we compute the third integral, $4\pi^2\lambda^2 J_3^2$, where
\begin{eqnarray}
 J_3 &=& \int \f{d^2 q}{(2\pi)^2} \f{(p+q)_{-}}{(p-q)_{-}}
\int \f{d^{(1-\epsilon)}\hat q_3}{(2 \pi)^{1-\epsilon}} \f{1}{q_s^2+\hat q_3^2}\nonumber \\
&=&A \int \f{d^2 q}{(2\pi)^2} \f{(p+q)_{-}}{(p-q)_{-}} \f{1}{|q_s|^{1+\epsilon}} \nn
&=&A \times\left({1 \over \pi(1-\epsilon) |p_s|^{\epsilon-1} } \right) \nn
&=& \f{p}{2 \pi} +\mathcal{O}(\epsilon).
\end{eqnarray}
Since $2 J_2+J_3^2=0$ with $\epsilon\to0$ by analytic continuation, the second and third terms in (\ref{DSeq0}) cancel. 

Finally we turn to the fourth term. 
By using the solution \eqref{solutionfermion}, the fourth term becomes
$-2 N \lambda_{4} f_0 J_4$, where 
\be
J_4 =  \int \f{d^3q}{(2\pi)^3} \f{q_s }{q_s^2+q_3^2}.
\label{j4}
\ee
Following the prescription in Appendix \ref{regularization} we regularize this integral as follows. 
\be
J_4 =  \int \f{d^2 q}{(2\pi)^2} 
\int \f{d^{(1-\epsilon)}\hat q_3}{(2 \pi)^{1-\epsilon}} \f{\sqrt{q_s^2+m^2} }{q_s^2+m^2+\hat q_3^2},
\ee
where we insert a virtual mass in 2-plane to cure IR divergence. 
Performing integration we find
\beal{
J_4 &= A \int \f{d^2 q}{(2\pi)^2}  \f{\sqrt{q_s^2+m^2} }{\sqrt{q_s^2+m^2}^{1+\epsilon}}
=-A {m^{2-\epsilon} \over 2\pi (2-\epsilon)} = - {m^2 \over 8\pi} + \cO(\epsilon),
}
which vanishes under $m\to0, \epsilon\to0$.

As a result we complete checking the equation \eqref{DSeq0}.

\subsection{Integration formulas}
\label{formulas}

In this appendix we collect detailed calculation used in this paper.

We first show the following integration formula.
\beal{
\frac{1}{\beta}\sum_{n} 
\int_{p}^\infty \frac{q_s dq_s}{(2\pi)} \frac{1}{ (\frac{2 \pi n}{\beta})^2+ q_s^2  + M^2 }
&=- { 1 \over 2\pi \beta} \log\left(2 \sinh ( \frac{\beta\sqrt{p^2 + M^2}}{2}) \right). 
\label{form1}
}
We regularize the left-hand side such that 
\be
\mbox{(l.h.s. in \eqref{form1})} =\frac{1}{\beta}\sum_{n} 
\int_{p_s}^\infty \frac{q_s dq_s}{(2\pi)} \int\frac{d^{-\epsilon} \hat q}{(2\pi)^{-\epsilon}} \frac{1}{ (\frac{2 \pi n}{\beta})^2+ \hat q^2+ q_s^2  + M^2 }. 
\ee
By using the formula 
\be
{1 \over \beta} \sum_{n \in \bZ}  \frac{1}{ (\frac{2 \pi n}{\beta})^2+ h^2 }
={\coth ({\beta h \over 2}) \over 2 h},
\label{sumform1}
\ee
where $h$ is some positive constant, 
we can perform the summation.
\beal{ 
\mbox{(l.h.s. in \eqref{form1})} &=
\int_{p}^\infty \frac{q_s dq_s}{(2\pi)} \int\frac{d^{-\epsilon} \hat q}{(2\pi)^{-\epsilon}} 
\frac{\coth ({\beta \sqrt{ \hat q^2+ q_s^2  + M^2 }\over 2})}{ 2 \sqrt{ \hat q^2+ q_s^2  + M^2 }} \nn
&=
\int_{p}^\infty \frac{q_s dq_s}{(2\pi)} \int\frac{d^{-\epsilon} \hat q}{(2\pi)^{-\epsilon}} 
\biggl(\frac{\coth ({\beta \sqrt{ \hat q^2+ q_s^2  + M^2 }\over 2}) -1 }{ 2 \sqrt{ \hat q^2+ q_s^2  + M^2 }} + \frac{1 }{ 2 \sqrt{ \hat q^2+ q_s^2  + M^2 }}\biggr). 
}
In the second equality, we separate the divergent part as the second term so that 
the first term is convergent. 
Therefore the first term can be evaluated without calculating the integral of extra dimension 
and results in 
\be
\int_{p}^\infty \frac{q_s dq_s}{(2\pi)} 
\frac{\coth ({\beta \sqrt{ q_s^2  + M^2 }\over 2}) -1 }{ 2 \sqrt{ q_s^2  + M^2 }}
= {- 2 \log \left( 2 \sinh ( \frac{\beta\sqrt{p^2 + M^2}}{2})\right) +\beta\sqrt{p^2 + M^2}\over 4\pi \beta}.
\ee
The second term is calculated as
\beal{
\int_{p}^\infty \frac{q_s dq_s}{(2\pi)} \int\frac{d^{-\epsilon} \hat q}{(2\pi)^{-\epsilon}} 
 \frac{1 }{ 2 \sqrt{ \hat q^2+ q_s^2  + M^2 }}
 &= \half\int_{p}^\infty \frac{q_s dq_s}{(2\pi)}{1\over (4 \pi)^{- \epsilon\over 2}} {\Gamma(\half+{ \epsilon\over 2}) \over \Gamma(\half) } (q_s^2  + M^2)^{-\half-{ \epsilon\over 2}}\nn
 &=\half {1\over (4 \pi)^{- \epsilon\over 2}} {\Gamma(\half+{ \epsilon\over 2}) \over \Gamma(\half) } \left( -{(p^2  + M^2)^{1-\epsilon \over 2} \over 2\pi(1 -\epsilon)}\right).
 }
Note that  $\epsilon$ has to be greater than one to make this integral convergent. 
By analytic continuation from some positive number $\epsilon$ to zero, 
we finish evaluating the second term as $-{\sqrt{p^2+M^2} \over 4\pi}$.
Summing up these we obtain \eqref{form1}.

By setting $p=0$ for \eqref{form1} we obtain 
\beal{
\frac{1}{\beta}\sum_{n} 
\int \frac{d^2 q}{(2\pi)^2} \frac{1}{ (\frac{2 \pi n}{\beta})^2+ q_s^2  + M^2 }
&=- { 1 \over 2\pi \beta} \log\left(2 \sinh ( \frac{\beta |M|}{2}) \right). 
\label{form1p0}
}

We can obtain a similar formula to \eqref{form1p0} by changing it from integer sum to half-integer sum. 
\beal{
\frac{1}{\beta}\sum_{n} 
\int \frac{d^2 q}{(2\pi)^2} \frac{1}{ (\frac{2 \pi (n+\half)}{\beta})^2+ q_s^2  + M^2 }
&=- { 1 \over 2\pi \beta} \log\left(2 \cosh ( \frac{\beta |M|}{2}) \right). 
\label{form2}
}
One can show this formula by repeating the same calculation 
by exchanging $\sinh X, \coth X$ into $\cosh X, \tanh X$, respectively.
This is because 
the summation formula \eqref{sumform1} is correspondingly changed by
\be
{1 \over \beta} \sum_{n \in \bZ}  \frac{1}{ (\frac{2 \pi (n+\half)}{\beta})^2+ h^2 }
={\tanh ({\beta h \over 2}) \over 2 h}.
\label{sumform2}
\ee

\bigskip

Next we generalize the formula \eqref{form2} in the following manner. 
\be
\frac{1}{\beta} \sum_{n}\int \frac{d^2 q}{(2\pi)^2} 
 \f{\left(\log\left(2\cosh({\beta\sqrt{q_s^2+M^2}\over 2})\right)\right)^a }{(\f{2 \pi (n+\half)}{\beta})^2+q_s^2+M^2}=
 -{ \left(\log\left(2\cosh({\beta M \over 2})\right) \right)^{a+1}    \over 2\pi \beta(a+1)}. 
\label{form3}
\ee
The case with $a=0$ is obtained above.
The procedure to calculate is the same above. 
We regularize the left-hand side such that 
\be
\mbox{(l.h.s. in \eqref{form3})} =\frac{1}{\beta}\sum_{n} 
\int \frac{d^2q}{(2\pi)^2} \int\frac{d^{-\epsilon} \hat q}{(2\pi)^{-\epsilon}} \frac{\left(\log\left(2\cosh({\beta\sqrt{q_s^2+M^2}\over 2})\right)\right)^a }{ (\frac{2 \pi n}{\beta})^2+ \hat q^2+ q_s^2  + M^2}. 
\ee
Employing the formula \eqref{sumform1} 
we can carry out the sum
\beal{ 
\mbox{(l.h.s. in \eqref{form3})} &=
\int \frac{d^2q}{(2\pi)^2} \int\frac{d^{-\epsilon} \hat q}{(2\pi)^{-\epsilon}} 
\frac{\left(\log\left(2\cosh({\beta\sqrt{q_s^2+M^2}\over 2})\right)\right)^a \coth ({\beta \sqrt{ \hat q^2+ q_s^2  + M^2 }\over 2})}{ 2 \sqrt{ \hat q^2+ q_s^2  +M^2 }} \nn
&=
\int \frac{ d^2q}{(2\pi)^2} \int\frac{d^{-\epsilon} \hat q}{(2\pi)^{-\epsilon}} 
\frac{\left(\log\left(2\cosh({\beta\sqrt{q_s^2+M^2}\over 2})\right)\right)^a \coth ({\beta \sqrt{ \hat q^2+ q_s^2  + M^2 }\over 2}) -  \left({\beta\sqrt{q_s^2+M^2}\over 2}\right)^a} { 2 \sqrt{ \hat q^2+ q_s^2  + M^2 }}  \nn
&\qquad +\int \frac{ d^2q}{(2\pi)^2} \int\frac{d^{-\epsilon} \hat q}{(2\pi)^{-\epsilon}}  \frac{\left({\beta\sqrt{q_s^2+M^2}\over 2}\right)^a}{ 2 \sqrt{ \hat q^2+ q_s^2  + M^2 }}. 
}
In the second equality we separate the divergent part. 
Now the first term is convergent, so it can be evaluated without taking into account the integral of extra dimension as follows. 
\beal{
&\int \frac{ d^2q}{(2\pi)^2} 
\frac{\left(\log\left(2\cosh({\beta\sqrt{q_s^2+M^2}\over 2})\right)\right)^a\coth ({\beta \sqrt{ q_s^2  + M^2 }\over 2}) -  \left({\beta\sqrt{q_s^2+M^2}\over 2}\right)^a }{ 2 \sqrt{q_s^2  + M^2 }} \nn
&= {- \left(\log\left(2\cosh({\beta M \over 2})\right) \right)^{a+1}    \over 2\pi \beta (a+1) } 
+ {1 \over 4\pi} \left({ \beta \over 2}\right)^a { M^{a+1} \over a+1}.
}
The second term is calculated as
\beal{
\half \left({\beta \over 2}\right)^a \int \frac{ d^2q}{(2\pi)^2} \int\frac{d^{-\epsilon} \hat q}{(2\pi)^{-\epsilon}}  \frac{({\sqrt{q_s^2+M^2}})^a}{ \sqrt{ \hat q^2+ q_s^2  + M^2 }} 
&=\half \left({\beta \over 2}\right)^a\int \frac{ d^2q}{(2\pi)^2}  {1\over (4 \pi)^{- \epsilon\over 2}} {\Gamma(\half+{ \epsilon\over 2}) \over \Gamma(\half) } (q_s^2  + M^2)^{{a\over 2} -\half-{ \epsilon\over 2}}\nn
&=\half \left({\beta \over 2}\right)^a{1\over (4 \pi)^{- \epsilon\over 2}} {\Gamma(\half+{ \epsilon\over 2}) \over \Gamma(\half) } \left( -{M^{a+1- \epsilon } \over 2\pi(a+1 -\epsilon)}\right).
 }
Note that  $\epsilon$ has to be greater than $a+1$ to make this integral convergent. 
By analytic continuation from some positive number $\epsilon$ to zero, 
we finish evaluating the second term as $-{1 \over 4\pi} \left({ \beta \over 2}\right)^a { M^{a+1} \over a+1}$.
Summing up these we obtain \eqref{form3}.

\section{Diagrammatic analysis}
\label{diagram}
In this appendix 
we check the gap equations (\ref{gapeqnall}) and the exact effective action (\ref{finansfe}) from a diagrammatic point of view.

First we consider the one-particle irreducible (1PI) diagram for bosonic self energy.%
\footnote{Remind that a 1PI diagram is such that it cannot be split into two by cutting any line from it.}
The bosonic 1PI self energy diagram is drawn in Fig.\ref{selfenergyboson}.
 \begin{figure}[t]
 \centerline{\includegraphics[scale=.9]{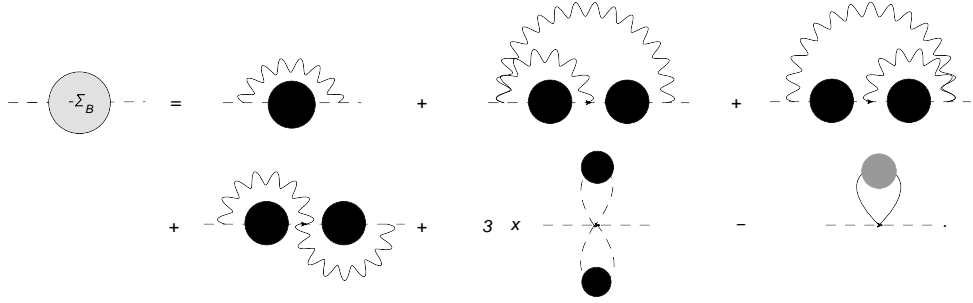}}
 \caption{ The bosonic 1PI self energy diagram is drawn. The dotted, undotted and wavy lines respectively represent scalar, fermion and gauge field. The black and gray bubble respectively stand for the planar propagators of scalar and fermion. }
 \label{selfenergyboson}
 \end{figure}
Note that gauge self interaction is not involved in these diagrams due to our gauge choice. 
This diagram encodes an equation such that\footnote{Note that 1PI self energy diagrams are given by $-\Sigma_B, -\Sigma_F$ in our notation. } 
\beal{
- \Sigma_B(p) &= N \int \frac{d^3q}{(2\pi)^3} 
(p+q)^\mu G_{\mu\nu}(p-q) \frac{1}{q^2+\Sigma_B(q)} (q+p)^\nu \nonumber \\
&- 2\times N^2 \int \frac{d^3q}{(2\pi)^3}\frac{d^3q'}{(2\pi)^3}
\frac{1}{q'^2+\Sigma_B(q')} G_{\mu\nu}(q-q') (q'+q)^\nu 
\frac{1}{q^2+\Sigma_B(q)} G^\mu{}_\rho(p-q) (p+q)^\rho
\nonumber \\
&- N^2 \int \frac{d^3q}{(2\pi)^3}\frac{d^3q'}{(2\pi)^3}
 (p+q')^\nu \frac{1}{q'^2+\Sigma_B(q')} G_{\nu\mu}(p-q')
\frac{1}{q^2+\Sigma_B(q)} G^\mu{}_\rho(p-q) (p+q)^\rho \nn
&-3\lambda_6 N^2 \int \frac{d^3q}{(2\pi)^3}\frac{d^3q'}{(2\pi)^3}\frac{1}{q^2+\Sigma_B(q)}\frac{1}{q'^2+\Sigma_B(q')}\nn
&-\lambda_4 N\int \frac{d^3q}{(2\pi)^3} \left(-\tr \frac{1 }{i\gamma^\mu q_\mu+\Sigma_F(q)} \right),
\label{FDboson}
}
where 
$G_{\mu\nu}$ is the relevant coefficient of the gauge propagator%
\footnote{ The gauge propagator is given by 
$
\aver{A^a_\mu(-p')  A^b_\nu(p)}
= \delta^{ab} (2 \pi)^3 \delta^3 (p-p') G_{\mu\nu}(p).$
}
 given by
\bea
G_{\mu\nu}(p)= {2\pi \over kip_- } (\delta_{\mu, +} \delta_{\nu, 3} - \delta_{\mu, 3} \delta_{\nu, +}).
\eea
One can easily reproduce the gap equation of boson in \eqref{gapeqnall} from this equation\footnote{We note
 that the first term in \eqref{FDboson} vanishes.}.

Next 
the fermionic 1PI self energy diagram is shown in Fig.\ref{selfenergyfermion}.
 \begin{figure}[t]
 \centerline{\includegraphics[scale=.9]{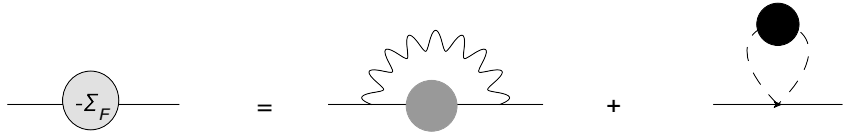}}
 \caption{ The fermionic 1PI self energy diagram is drawn.}
 \label{selfenergyfermion}
 \end{figure}
We can read off an equation from this diagram such that 
\beal{
- \Sigma_F(p) &= N \int \frac{d^3q}{(2\pi)^3} 
i\gamma^\mu G_{\mu\nu}(p-q) \frac{1 }{i\gamma^\mu q_\mu+\Sigma_F(q)} i\gamma^\nu - \lambda_4 N\int \frac{d^3q}{(2\pi)^3} \frac{1}{q^2+\Sigma_B(q)}.
\label{FDfermion}
}
It turns out that this agrees with the gap equation of fermion in \eqref{gapeqnall}.

Finally we check that 
the effective action (\ref{finansfe}) encodes the connected vacuum graph in a perturbative way. 
For this purpose, 
we first expand $-S_{eff}$ (\ref{finansfe}) by $-\Sigma_B, -\Sigma_F$, 
which is diagrammatically drawn in Fig.\ref{vacuumgraph}.
 \begin{figure}[h]
 \centerline{\includegraphics[scale=1]{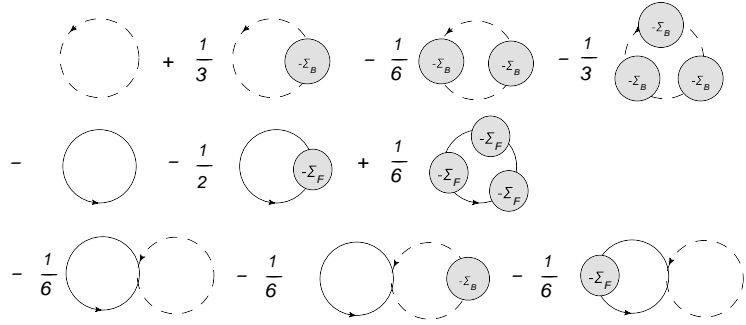}}
 \caption{We draw the perturbative expansion of $-S_{eff}$ in terms of $-\Sigma_B, -\Sigma_F$ diagrammatically, which is expected to describe the connected vacuum graph.}
 \label{vacuumgraph}
 \end{figure} 
Plugging \eqref{FDboson} and \eqref{FDfermion} into the equation expanded by $-\Sigma_B, -\Sigma_F$, 
we obtain the perturbative expansion in terms of $\lambda, x_4, x_6$. 
Each term represents a planar diagram contributing to the vacuum graph. 
These planar diagrams are categorized into three classes by matter content: 
One includes only scalar, another only fermion, the other both. 
The first class, which include only scalar and gauge field propagators, is the following.%
\footnote{If we normalize $\lambda_6\to{\lambda_6 \over 3}$, 
then the 6th and 7th terms respectively have the coefficients ${1\over 3}$ and  $3$.
}
\vspace{-.5cm}
\begin{center}
\begin{fmffile}{boson}
        \begin{tabular}{c}
            \begin{fmfgraph*}(10,10)
                \fmfsurroundn{w}{2}
                \fmf{dashes,tension=0,right=1}{w1,w2,w1}
                \end{fmfgraph*}
        \end{tabular}
$+$
        \begin{tabular}{c}
            \begin{fmfgraph*}(10,10)
                \fmfsurroundn{w}{6}
                \fmf{dashes,tension=0,right=1}{w1,w4,w1}
                \fmf{wiggly,tension0}{w1,w3}
                \fmf{wiggly,tension0}{w1,w5}
                \end{fmfgraph*}
        \end{tabular}
$+2$
        \begin{tabular}{c}
            \begin{fmfgraph*}(10,10)
                \fmfsurroundn{w}{10}
                \fmf{dashes,tension=0,right=1}{w1,w6,w1}
                \fmf{wiggly,tension0}{w9,w2}
                \fmf{wiggly,tension0}{w9,w3}
                \fmf{wiggly,left=.3}{w4,w6}
                \fmf{wiggly,left=.3}{w6,w8 }
                \end{fmfgraph*}
        \end{tabular}
$+2$
        \begin{tabular}{c}
            \begin{fmfgraph*}(10,10)
                \fmfsurroundn{w}{10}
                \fmf{dashes,tension=0,right=1}{w1,w6,w1}
                \fmf{wiggly,tension0}{w9,w2}
                \fmf{wiggly,tension0}{w9,w3}
                \fmf{wiggly}{w8,w4}
                \fmf{wiggly}{w8,w5}
                \end{fmfgraph*}
        \end{tabular}
$+{1\over 2}$
        \begin{tabular}{c}
            \begin{fmfgraph*}(10,10)
                \fmfsurroundn{w}{10}
                \fmf{dashes,tension=0,right=1}{w1,w6,w1}
                \fmf{wiggly,left=.3}{w1,w3}
                \fmf{wiggly,left=.3}{w3,w5}
                \fmf{wiggly,left=.3}{w6,w8}
                \fmf{wiggly,left=.3}{w8,w10}
                \end{fmfgraph*}
        \end{tabular} 
$+\!\!\!\!\!\!\!\!$
	\begin{tabular}{c}
            \begin{fmfgraph}(20,20)
            	\fmfleft{i}
		\fmfright{o1,o2}
		\fmfdot{v}
		\fmf{phantom}{i,v,v,o2}
		\fmf{dashes,tension=1,left}{v,v}
		\fmf{dashes,tension=1}{v,v}
		\fmf{phantom}{o1,v,v,i}
            \end{fmfgraph}
         \end{tabular}
\hspace{-3.5cm}
         \begin{tabular}{c}
	  \rotatebox{240}{\begin{fmfgraph}(20,20)
            	\fmfleft{i}
		\fmfright{o1,o2}
		\fmf{phantom}{i,v,v,o2}
		\fmf{dashes,tension=1,left}{v,v}
		\fmf{phantom}{o1,v,v,i}
            \end{fmfgraph}}
         \end{tabular}
\vspace{-5mm}
$\!\! \!\!\!\!\!\!+9 \!\!\!\!\!\!\!\!$
	\begin{tabular}{c}
            \begin{fmfgraph}(20,20)
            	\fmfleft{i}
		\fmfright{o1,o2}
		\fmfdot{v}
		\fmf{phantom}{i,v,v,o2}
		\fmf{dashes,tension=1,left}{v,v}
		\fmf{dashes,tension=1}{v,v}
		\fmf{phantom}{o1,v,v,i}
            \end{fmfgraph}
         \end{tabular}
\hspace{-3.5cm}
         \begin{tabular}{c}
	  \rotatebox{240}{\begin{fmfgraph}(20,20)
            	\fmfleft{i}
		\fmfright{o1,o2}
		\fmf{phantom}{i,v,v,o2}
		\fmf{dashes,tension=1,left}{v,v}
		\fmf{phantom}{o1,v,v,i}
            \end{fmfgraph}}
         \end{tabular}
\hspace{-1.68cm}
        \begin{tabular}{c}
            \begin{fmfgraph*}(4,4)
                \fmfsurroundn{w}{10}
                \fmf{phantom,tension=0,right=1}{w1,w6,w1}
                \fmf{wiggly,tension0}{w1,w5}
                \fmf{wiggly,tension0}{w1,w7}
                \end{fmfgraph*}
        \end{tabular}
\end{fmffile}
$+\quad\cdots$.
\end{center}
The 2nd class, which involves only fermion and gauge propagators, is 
\begin{center}
$- \!\!\!\!$
\begin{fmffile}{fermion}
        \begin{tabular}{c}
            \begin{fmfgraph*}(15,10)
                \fmfsurroundn{w}{2}
                \fmf{plain,tension=0,right=1}{w1,w2,w1}
                \end{fmfgraph*}
        \end{tabular}
$\!\!\!\!\!\!-{1\over 2}$
        \begin{tabular}{c}
            \begin{fmfgraph*}(15,10)
                \fmfsurroundn{w}{2}
                \fmf{plain,tension=0,right=1}{w1,w2,w1}
                \fmf{wiggly,tension0}{w1,w2}
                \end{fmfgraph*}
        \end{tabular}
$\!\!\!\!\!\!-{1\over 2}$
        \begin{tabular}{c}
            \begin{fmfgraph*}(15,10)
                \fmfsurroundn{w}{4}
                \fmf{plain,tension=0,right=1}{w1,w3,w1}
                \fmf{wiggly,tension0}{w1,w2}
                \fmf{wiggly,tension0}{w3,w4}
                \end{fmfgraph*}
        \end{tabular}
$\!\!\!\!\!\!-{1\over 3}$
        \begin{tabular}{c}
            \begin{fmfgraph*}(15,10)
                \fmfsurroundn{w}{6}
                \fmf{plain,tension=0,right=1}{w1,w4,w1}
                \fmf{wiggly,left=.5}{w1,w2}
                \fmf{wiggly,left=.5}{w3,w4}
                \fmf{wiggly,left=.5}{w5,w6}
                \end{fmfgraph*}
        \end{tabular}
$\!\!\!\!\!\!-{1\over 2}$
        \begin{tabular}{c}
            \begin{fmfgraph*}(15,10)
                \fmfsurroundn{w}{6}
                \fmf{plain,tension=0,right=1}{w1,w4,w1}
                \fmf{wiggly,right=.5}{w1,w6}
                \fmf{wiggly}{w2,w5}
                \fmf{wiggly,left=.5}{w3,w4}
                \end{fmfgraph*}
        \end{tabular}
        \end{fmffile}
$\!\!\!\!\!+\quad \cdots$.
\end{center}
The third class which includes both scalar and fermion lines is given by
\begin{center}
\begin{fmffile}{bosonfermion}
$-$
        \begin{tabular}{c}
            \begin{fmfgraph*}(15,10)
            	\fmfleft{i}
		\fmfright{o}
		\fmf{phantom,tension=10}{i,i1}
		\fmf{phantom,tension=10}{o,o1}
		\fmf{plain,tension=.4,left}{i1,v,i1}
		\fmf{dashes,tension=.4,right}{o1,v,o1}
                \end{fmfgraph*}
        \end{tabular}
$+{1\over 2}$
        \begin{tabular}{c}
            \begin{fmfgraph*}(20,20)
            	\fmfleft{i}
		\fmfright{o}
		\fmf{phantom,tension=10}{i,i1}
		\fmf{phantom,tension=10}{o,o1}
		\fmf{plain,tension=.4,left}{i1,v1,i1}
		\fmf{plain,tension=.4,right}{o1,v2,o1}
		\fmf{dashes,tension=.4,right}{v1,v2,v1}
                \end{fmfgraph*}
        \end{tabular}
$-{1\over 2}$
        \begin{tabular}{c}
            \begin{fmfgraph*}(20,20)
            	\fmfleft{i}
		\fmfright{o}
		\fmf{phantom,tension=10}{i,i1}
		\fmf{phantom,tension=10}{o,o1}
		\fmf{dashes,tension=.4,left}{i1,v1,i1}
		\fmf{dashes,tension=.4,right}{o1,v2,o1}
		\fmf{plain,tension=.4,right}{v1,v2,v1}
                \end{fmfgraph*}
        \end{tabular}
        \end{fmffile}
$+\quad \cdots$.
\end{center}

These contributions can be explicitly checked by perturbative calculation of the connected vacuum graph by using the original action \eqref{LCaction1} including gauge field. In fact, one can easily see that 
each diagram has the correct symmetry and sign factors.%
\footnote{
The graph including $n_F$ fermion loops has the sign factor $(-1)^{n_F}$.
}

\section{Supersymmetric Chern-Simons matter action}
\label{susycs}
In this appendix we present $\cN=1, 2$ Chern-Simons theory coupling to a 
fundamental chiral multiplet in our notation.

\subsection{$\cN=2$ case}
\label{susycsn2}
 
$\cN=2$ Chern-Simons-matter Lagrangian we use consists of
$\cN=2$ gauge multiplet $(A_\mu, \sigma, \lambda, \bar\lambda, D)$, and a chiral multiplet $(\phi, \psi, F)$. 
The Lagrangian is 
\cite{Ivanov:1991fn} 
\be
\cL^{\cN=2} =  \cL^{\cN=2}_{cs} + \cL^{\cN=2}_{matter},
\ee
where
\bea
 \cL^{\cN=2}_{cs}
&=&\kappa \Tr \left[i\varepsilon^{\mu\nu\rho}(A_\mu\partial_\nu A_\rho -{2i\over3}  A_\mu  A_\nu  A_\rho)
-\bar \lambda \lambda +2 D \sigma \right], \\
 \cL^{\cN=2}_{matter}
&=& D_\mu \bar \phi  D^\mu \phi +  \bar\psi (\sla D+\sigma)\psi 
+ \bar\psi \bar\lambda \phi + \bar \phi \lambda\psi + \bar \phi  (\sigma^2-D)\phi - \bar F  F.
\eea
Here $\varepsilon^{123}=1$, $\sla D = \gamma^\mu D_\mu$.
Supersymmetry transformation rule for a vector multipelet is 
\bea
\Delta _\epsilon  A_{\mu}&=&
{i\over \sqrt2} (\bar \epsilon \gamma_\mu \lambda - \epsilon \gamma_\mu \bar \lambda), \\
\Delta _\epsilon \sigma &=&-{1\over \sqrt2} (\epsilon \bar \lambda +\bar \epsilon \lambda),\\
\Delta _\epsilon \lambda_\alpha &=& 
 \sqrt2 \epsilon_\gamma \gamma_\rho{}^\gamma{}_\alpha({i\over 2} \varepsilon^{\mu\nu\rho}F_{\mu\nu}+[D^\rho,\sigma])-  \sqrt2 \epsilon_\alpha D,\\
\Delta _\epsilon \bar\lambda_\alpha &=& 
 \sqrt2 \bar\epsilon_\gamma \gamma_\rho{}^\gamma{}_\alpha(-{i\over 2} \varepsilon^{\mu\nu\rho}F_{\mu\nu}+[D^\rho,\sigma])-  \sqrt2\bar\epsilon_\alpha D,\\
\Delta _\epsilon D &=&
-{1\over \sqrt2}(\epsilon_\gamma \sla D^\gamma{}_\beta \bar\lambda^\beta+\bar\epsilon_\gamma \sla D^\gamma{}_\beta\lambda^\beta + \epsilon[\bar\lambda,\sigma]+\bar\epsilon[\sigma,\lambda]).
\eea
For a chiral multiplet, 
\bea
\Delta _\epsilon \phi&=&-\sqrt2  \epsilon\psi,\\
\Delta _\epsilon \psi_\alpha&=& 
\sqrt2 (\bar\epsilon_\beta\sla{D}^{\beta}{}_{\alpha}\phi+\bar\epsilon_{\alpha}\sigma \phi-\epsilon_\alpha F),\\
\Delta _\epsilon F&=& -\sqrt2(\bar\epsilon \sla{D} \psi+  \bar\epsilon\sigma \psi + \bar\epsilon \bar\lambda \phi), \\
\Delta _\epsilon \bar \phi &=&- \sqrt2 \bar\epsilon \bar \psi, \\
\Delta _\epsilon \bar \psi_\alpha &=& 
\sqrt2 (\epsilon_\beta \sla{D}^{\beta}{}_{\alpha}\bar \phi +\epsilon_{\alpha}\bar \phi \sigma
-\bar\epsilon_\alpha \bar F ),\\
\Delta _\epsilon \bar F &=&- \sqrt2( \epsilon \sla{ D}\bar \psi +\epsilon \bar \psi\sigma +  \bar \phi  \epsilon \lambda).
\eea

Integrating out $D, \sigma, \lambda, \bar\lambda, F$
we obtain $\cN=2$ Chern-Simons action 
\beal{
S^{\cN=2} 
&=\int d^3 x \biggl[ i\kappa \varepsilon^{\mu\nu\rho}\Tr (A_\mu\partial_\nu A_\rho -{2i\over3}  A_\mu  A_\nu A_\rho) + D_\mu \bar \phi D^\mu \phi + \bar\psi \sla D\psi \nn
& \qquad + {1 -\frac{2}{N} \over 2 \kappa} (\bar\psi \phi)( \bar \phi \psi)
+ {1 -\frac{1}{2N} \over \kappa} (\bar\psi \psi) (\bar \phi \phi) 
+ \left( {1-\frac{1}{N} \over 2 \kappa}\right)^2 (\bar \phi \phi)^3 \biggl],
}
and supersymmetry transformation rule 
\bea
\Delta _\epsilon A^a_{\mu}&=&
{i \over \sqrt2\kappa} (\phi\bar T^a\epsilon \gamma_\mu \bar\psi -\epsilon \gamma_\mu \psi T^a \bar \phi), \\
\Delta _\epsilon \phi&=&-\sqrt2  \epsilon\psi,\\
\Delta _\epsilon \psi_\alpha&=& 
\sqrt2 (\bar\epsilon_\beta \sla{D}^{\beta}{}_{\alpha} \phi 
+{1-{1\over N} \over 2\kappa} \bar\epsilon_\alpha \phi (\bar \phi \phi)),\\
\Delta _\epsilon \bar \phi &=&- \sqrt2 \bar\epsilon \bar \psi, \\
\Delta _\epsilon \bar \psi_\alpha &=& 
\sqrt2 (\epsilon_\beta  \sla{D}^{\beta}{}_{\alpha}\bar \phi 
+{1-{1\over N} \over 2\kappa}\epsilon_{\alpha}(\bar \phi \phi) \bar \phi ).
\eea


\subsection{$\cN=1$ case}
\label{susycsn1}

$\cN=1$ Chern-Simons-matter Lagrangian we use consists of
$\cN=1$ gauge multiplet $(A_\mu, \lambda')$, where $\lambda'$ is a majorana fermion, and matter multiplet $(\phi, \psi)$, where $\phi$ is a complex boson. 
The Lagrangian with a general superpotential $W(\bar\phi \phi)$ is given by
\cite{Lee:1990it}
\be
\cL^{{\cal N}=1} =  \cL^{{\cal N}=1}_{cs} + \cL^{{\cal N}=1}_{matter},
\ee
where
\bea
{\cal L}^{{\cal N}=1}_{cs}&=&{ \kappa } \Tr \left[i\varepsilon^{\mu\nu\rho}( A_\mu\partial_\nu  A_\rho -{2i\over3} A_\mu  A_\nu A_\rho) - \lambda'_\alpha \lambda'^\alpha \right], \\
{\cal L}^{{\cal N}=1}_{matter} &=& \biggl[D_\mu \bar \phi D^\mu \phi + \bar\psi \slash\!\!\!\! D \psi 
+i\bar\psi \lambda' \phi - i \bar \phi\lambda'\psi+ (\bar \phi \phi) W'({\bar \phi \phi})^2- \bar\psi \psi W'({\bar \phi \phi})  \nn
&&\quad  
-\biggl( \half \left((\bar\psi \phi) (\bar\psi \phi) + (\psi\bar \phi)( \psi \bar \phi)\right)
+ (\bar \phi \psi) (\bar \psi \phi)\biggl) W''({\bar \phi\phi}) \biggl].
\eea
Here $W'(x) = {d W(x) \over d x}=:W'_x$.  
Supersymmetry transformation rule is 
\bea
\Delta_\epsilon A_{\mu}&=& \sqrt 2 \epsilon \gamma_{\mu} \lambda',\\
\Delta_\epsilon \lambda'_\alpha &=& -{1\over \sqrt 2} \varepsilon^{\mu\nu\rho}
\epsilon_\beta \gamma_\rho{}^\beta{}_\alpha F_{\mu\nu},\\
\Delta_\epsilon \phi&=& -\sqrt 2 \epsilon\psi,\\
\Delta_\epsilon \psi_\alpha&=& 
\sqrt 2 (\epsilon_\beta \sla D^{\beta}{}_{\alpha}\phi-\epsilon_\alpha \phi W'(\bar \phi \phi)),\\
\Delta_\epsilon \bar \phi&=& -\sqrt 2 \epsilon\bar \psi,\\
\Delta_\epsilon \bar\psi_\alpha&=& 
\sqrt 2 (\epsilon_\beta \sla D^{\beta}{}_{\alpha}\bar \phi-\epsilon_\alpha W'(\bar \phi \phi) \bar \phi).
\eea

Integrating out gaugino $\lambda'$ we obtain $\cN=1$ Chern-Simons action
\beal{
S^{{\cal N}=1}  
&= \int d^3 x \biggl[ i\kappa\varepsilon^{\mu\nu\rho}\Tr ( A_\mu\partial_\nu A_\rho -{2i\over3} A_\mu A_\nu A_\rho)  
 + D_\mu \bar \phi D^\mu \phi + \bar\psi \slash\!\!\!\! D \psi \nn
&\qquad + (\bar \phi \phi) W'^2_{\bar \phi \phi}+ (\bar\psi \psi) \left( - W'_{\bar \phi \phi}+{1  \over 2\kappa} (\bar \phi \phi)\right) -(\bar \phi \psi) (\bar \psi \phi) (W''_{\bar \phi\phi} +{1 \over 2\kappa N})\nn
&\qquad  
+\left(- \half{W''_{\bar \phi \phi}} - {1-\frac{1}{N} \over 4 \kappa}\right) ((\bar\psi \phi)( \bar \psi \phi ) +(\bar \phi \psi)( \bar \phi \psi )) \biggl],
}
and supersymmetry transformation rule as
\bea
\Delta _\epsilon A_{\mu}^a&=&
{i \over \sqrt 2 \kappa} (\phi T^a \epsilon \gamma_\mu \bar\psi -\epsilon \gamma_\mu  \psi T^a\bar \phi), \\
\Delta_\epsilon \phi&=& -\sqrt 2 \epsilon\psi,\\
\Delta_\epsilon \psi_\alpha&=& 
\sqrt 2 (\epsilon_\beta \sla D^{\beta}{}_{\alpha}\phi -\epsilon_\alpha \phi W'(\bar \phi \phi)),\\
\Delta_\epsilon \bar \phi&=& -\sqrt 2 \epsilon\bar \psi,\\
\Delta_\epsilon \bar\psi_\alpha&=& 
\sqrt 2 (\epsilon_\beta \sla D^{\beta}{}_{\alpha}\bar \phi-\epsilon_\alpha W'(\bar \phi \phi) \bar \phi).
\eea

Now, let us impose the conformal symmetry on this action, 
which requires the potential $W(\bar \phi \phi)$ to be quadratic.
\be
W(\bar \phi \phi) = -{w \over 4 \kappa} (\bar \phi \phi)^2,
\label{superpot}
\ee
where $w$ is an arbitrary number. Then the above action becomes
\beal{
S^{{\cal N}=1}  
&= \int d^3 x \biggl[ i\kappa\varepsilon^{\mu\nu\rho}\Tr ( A_\mu\partial_\nu A_\rho -{2i\over3} A_\mu A_\nu A_\rho)  + D_\mu \bar \phi D^\mu \phi + \bar\psi \slash\!\!\!\! D \psi \nn
&\qquad + \left({w \over 2 \kappa}\right)^2(\bar \phi \phi)^3+  \left( {1  +w  \over 2\kappa} \right)(\bar\psi \psi)(\bar \phi \phi)  \nn
&\qquad
+{w -\frac{1}{N} \over 2 \kappa}(\bar \phi \psi) (\bar \psi \phi) 
+\left( { w-1+\frac{1}{N} \over 4 \kappa}\right) ((\bar\psi \phi)( \bar \psi \phi ) +(\bar \phi \psi)( \bar \phi \psi )) \biggl],
}
and supersymmetry transformation rule 
\bea
\Delta _\epsilon A_{\mu}^a&=&
{i \over \sqrt 2 \kappa} (\phi T^a \epsilon \gamma_\mu \bar\psi -\epsilon \gamma_\mu  \psi T^a\bar \phi), \\
\Delta_\epsilon \phi&=& -\sqrt 2 \epsilon\psi,\\
\Delta_\epsilon \psi_\alpha&=& 
\sqrt 2 (\epsilon_\beta \sla D^{\beta}{}_{\alpha}\phi +{w \over 2 \kappa} \epsilon_\alpha \phi (\bar \phi \phi)),\\
\Delta_\epsilon \bar \phi&=& -\sqrt 2 \epsilon\bar \psi,\\
\Delta_\epsilon \bar\psi_\alpha&=& 
\sqrt 2 (\epsilon_\beta \sla D^{\beta}{}_{\alpha}\bar \phi +{w \over 2 \kappa}\epsilon_\alpha (\bar \phi \phi) \bar \phi).
\eea

Note that $\cN=1$ SUSY in this theory is enhanced to $\cN=2$ when $w=1 - {1 \over N}$. This can be explicitly seen by comparing $\cN=2$ SUSY action constructed in Appendix \ref{susycsn2}.

\bibliographystyle{utphys}
\bibliography{refcssusyv2}

\providecommand{\href}[2]{#2}\begingroup\raggedright\begin{thebibliography}{10}

\bibitem{Maldacena:1997re}
J.~M. Maldacena, ``{The Large N limit of superconformal field theories and
  supergravity},'' {\em Adv.Theor.Math.Phys.} {\bf 2} (1998) 231--252,
\href{http://arXiv.org/abs/hep-th/9711200}{{\tt hep-th/9711200}}.

\bibitem{Strominger:2001pn}
A.~Strominger, ``{The dS / CFT correspondence},'' {\em JHEP} {\bf 0110} (2001)
  034,
\href{http://arXiv.org/abs/hep-th/0106113}{{\tt hep-th/0106113}}.

\bibitem{Klebanov:2002ja}
I.~Klebanov and A.~Polyakov, ``{AdS dual of the critical O(N) vector model},''
  {\em Phys.Lett.} {\bf B550} (2002) 213--219,
\href{http://arXiv.org/abs/hep-th/0210114}{{\tt hep-th/0210114}}.

\bibitem{Vasiliev:1990en}
M.~A. Vasiliev, ``{Consistent equation for interacting gauge fields of all
  spins in (3+1)-dimensions},'' {\em Phys.Lett.} {\bf B243} (1990)
378--382.

\bibitem{Vasiliev:2003ev}
M.~Vasiliev, ``{Nonlinear equations for symmetric massless higher spin fields
  in (A)dS(d)},'' {\em Phys.Lett.} {\bf B567} (2003) 139--151,
\href{http://arXiv.org/abs/hep-th/0304049}{{\tt hep-th/0304049}}.

\bibitem{Bekaert:2005vh}
X.~Bekaert, S.~Cnockaert, C.~Iazeolla, and M.~Vasiliev, ``{Nonlinear higher
  spin theories in various dimensions},''
\href{http://arXiv.org/abs/hep-th/0503128}{{\tt hep-th/0503128}}.

\bibitem{Sezgin:2002rt}
E.~Sezgin and P.~Sundell, ``{Massless higher spins and holography},'' {\em
  Nucl.Phys.} {\bf B644} (2002) 303--370,
\href{http://arXiv.org/abs/hep-th/0205131}{{\tt hep-th/0205131}}.

\bibitem{Sezgin:2003pt}
E.~Sezgin and P.~Sundell, ``{Holography in 4D (super) higher spin theories and
  a test via cubic scalar couplings},'' {\em JHEP} {\bf 0507} (2005) 044,
\href{http://arXiv.org/abs/hep-th/0305040}{{\tt hep-th/0305040}}.

\bibitem{Giombi:2010vg}
S.~Giombi and X.~Yin, ``{Higher Spins in AdS and Twistorial Holography},'' {\em
  JHEP} {\bf 1104} (2011) 086,
\href{http://arXiv.org/abs/1004.3736}{{\tt 1004.3736}}.

\bibitem{Giombi:2011rz}
S.~Giombi, S.~Prakash, and X.~Yin, ``{A Note on CFT Correlators in Three
  Dimensions},''
\href{http://arXiv.org/abs/1104.4317}{{\tt 1104.4317}}.

\bibitem{Giombi:2011ya}
S.~Giombi and X.~Yin, ``{On Higher Spin Gauge Theory and the Critical O(N)
  Model},''
\href{http://arXiv.org/abs/1105.4011}{{\tt 1105.4011}}.

\bibitem{Giombi:2011kc}
S.~Giombi, S.~Minwalla, S.~Prakash, S.~P. Trivedi, S.~R. Wadia, {\em et al.},
  ``{Chern-Simons Theory with Vector Fermion Matter},''
\href{http://arXiv.org/abs/1110.4386}{{\tt 1110.4386}}.

\bibitem{Aharony:2011jz}
O.~Aharony, G.~Gur-Ari, and R.~Yacoby, ``{d=3 Bosonic Vector Models Coupled to
  Chern-Simons Gauge Theories},'' {\em JHEP} {\bf 1203} (2012) 037,
\href{http://arXiv.org/abs/1110.4382}{{\tt 1110.4382}}.

\bibitem{Chang:2012kt}
C.-M. Chang, S.~Minwalla, T.~Sharma, and X.~Yin, ``{ABJ Triality: from Higher
  Spin Fields to Strings},''
\href{http://arXiv.org/abs/1207.4485}{{\tt 1207.4485}}.

\bibitem{Shenker:2011zf}
S.~H. Shenker and X.~Yin, ``{Vector Models in the Singlet Sector at Finite
  Temperature},''
\href{http://arXiv.org/abs/1109.3519}{{\tt 1109.3519}}.

\bibitem{Hellerman:2012}
S.~Banerjee, S.~Hellerman, J.~Maltz, and S.~H. Shenker, ``{Light States in
  Chern-Simons Theory Coupled to Fundamental Matter},''
\href{http://arXiv.org/abs/1207.4195}{{\tt 1207.4195}}.

\bibitem{Maldacena:2011jn}
J.~Maldacena and A.~Zhiboedov, ``{Constraining Conformal Field Theories with A
  Higher Spin Symmetry},''
\href{http://arXiv.org/abs/1112.1016}{{\tt 1112.1016}}.

\bibitem{Maldacena:2012sf}
J.~Maldacena and A.~Zhiboedov, ``{Constraining conformal field theories with a
  slightly broken higher spin symmetry},''
\href{http://arXiv.org/abs/1204.3882}{{\tt 1204.3882}}.

\bibitem{Gaiotto:2007qi}
D.~Gaiotto and X.~Yin, ``{Notes on superconformal Chern-Simons-Matter
  theories},'' {\em JHEP} {\bf 0708} (2007) 056,
\href{http://arXiv.org/abs/0704.3740}{{\tt 0704.3740}}.

\bibitem{Hubbard:1959ub}
J.~Hubbard, ``{Calculation of partition functions},'' {\em Phys.Rev.Lett.} {\bf
  3} (1959)
77--80.

\bibitem{Stratonovich:1959}
R.~L. Stratonovich, ``{On a Method of Calculating Quantum Distribution
  Functions},'' {\em Soviet Physics Doklady} {\bf 2} (1959)
416.

\bibitem{Aharony:2012nh}
O.~Aharony, G.~Gur-Ari, and R.~Yacoby, ``{Correlation Functions of Large N
  Chern-Simons-Matter Theories and Bosonization in Three Dimensions},''
\href{http://arXiv.org/abs/1207.4593}{{\tt 1207.4593}}.

\bibitem{Oferb:2012}
O.~Aharony, G.~Gur-Ari, and R.~Yacoby {\em unpublished}.

\bibitem{AlvarezGaume:2006jg}
L.~Alvarez-Gaume, P.~Basu, M.~Marino, and S.~R. Wadia, ``{Blackhole/String
  Transition for the Small Schwarzschild Blackhole of AdS(5)x S**5 and Critical
  Unitary Matrix Models},'' {\em Eur.Phys.J.} {\bf C48} (2006) 647--665,
\href{http://arXiv.org/abs/hep-th/0605041}{{\tt hep-th/0605041}}.

\bibitem{Avdeev:1992jt}
L.~Avdeev, D.~Kazakov, and I.~Kondrashuk, ``{Renormalizations in supersymmetric
  and nonsupersymmetric nonAbelian Chern-Simons field theories with matter},''
  {\em Nucl.Phys.} {\bf B391} (1993)
333--357.

\bibitem{Pisarski:1985yj}
R.~D. Pisarski and S.~Rao, ``{Topologically Massive Chromodynamics in the
  Perturbative Regime},'' {\em Phys.Rev.} {\bf D32} (1985)
2081.

\bibitem{Aharony:2008gk}
O.~Aharony, O.~Bergman, and D.~L. Jafferis, ``{Fractional M2-branes},'' {\em
  JHEP} {\bf 0811} (2008) 043,
\href{http://arXiv.org/abs/0807.4924}{{\tt 0807.4924}}.

\bibitem{Giveon:2008zn}
A.~Giveon and D.~Kutasov, ``{Seiberg Duality in Chern-Simons Theory},'' {\em
  Nucl. Phys.} {\bf B812} (2009) 1--11,
\href{http://arXiv.org/abs/0808.0360}{{\tt 0808.0360}}.

\bibitem{Gross:1980he}
D.~Gross and E.~Witten, ``{Possible Third Order Phase Transition in the Large N
  Lattice Gauge Theory},'' {\em Phys.Rev.} {\bf D21} (1980)
446--453.

\bibitem{Wadia:1980cp}
S.~R. Wadia, ``{N = infinity phase transition in a class of exactly soluable
  model lattice gauge theories},'' {\em Phys.Lett.} {\bf B93} (1980)
403.

\bibitem{Anninos:2011ui}
D.~Anninos, T.~Hartman, and A.~Strominger, ``{Higher Spin Realization of the
  dS/CFT Correspondence},''
\href{http://arXiv.org/abs/1108.5735}{{\tt 1108.5735}}.

\bibitem{Ng:2012xp}
G.~S. Ng and A.~Strominger, ``{State/Operator Correspondence in Higher-Spin
  dS/CFT},''
\href{http://arXiv.org/abs/1204.1057}{{\tt 1204.1057}}.

\bibitem{Das:2012dt}
D.~Das, S.~R. Das, A.~Jevicki, and Q.~Ye, ``{Bi-local Construction of Sp(2N)/dS
  Higher Spin Correspondence},''
\href{http://arXiv.org/abs/1205.5776}{{\tt 1205.5776}}.

\bibitem{Wadia:1980rb}
S.~R. Wadia, ``{On the Dyson-Schwinger equations approach to the large N limit:
  model systems and string representation of Yang-Mills theory},'' {\em
  Phys.Rev.} {\bf D24} (1981)
970.

\bibitem{Ivanov:1991fn}
E.~Ivanov, ``{Chern-Simons matter systems with manifest N=2 supersymmetry},''
  {\em Phys.Lett.} {\bf B268} (1991)
203--208.

\bibitem{Lee:1990it}
C.-k. Lee, K.-M. Lee, and E.~J. Weinberg, ``{Supersymmetry and selfdual
  Chern-Simons systems},'' {\em Phys.Lett.} {\bf B243} (1990)
105--108.

\end{thebibliography}\endgroup

\end{document}